\documentclass[prl,superscriptaddress,showpacs,twocolumn,longbibliography]{revtex4-1}

\usepackage{hyperref}
\usepackage{braket}
\usepackage{color}
\usepackage[usenames,dvipsnames]{xcolor}
\usepackage{amsmath,amsthm,amssymb}
\usepackage{graphicx}
\usepackage{epsfig}
\usepackage{dcolumn}
\usepackage{bm}
\usepackage{mathrsfs}
\usepackage{multirow}
\usepackage[all]{xy}
\usepackage{pbox}
\usepackage{verbatim}

\def\(({\left(}
\def\)){\right)}
\def\[[{\left[}
\def\]]{\right]}

\newcommand{\be}{\begin{equation}}
\newcommand{\ee}{\end{equation}}
\newcommand{\ben}{\begin{eqnarray}}
\newcommand{\een}{\end{eqnarray}}
\newcommand{\beq}{\begin{equation}}
\newcommand{\eeq}{\end{equation}}

\begin{document}

\title{Nonequilibrium many-body quantum engine driven by time-translation symmetry breaking}
\author{Federico Carollo}
\affiliation{Institut f\"ur Theoretische Physik, Universit\"at T\"ubingen, Auf der Morgenstelle 14, 72076 T\"ubingen, Germany}
\author{Kay Brandner}
\affiliation{School of Physics and Astronomy and \\Centre for the Mathematics and Theoretical Physics of Quantum Non-Equilibrium Systems, University of Nottingham, Nottingham, NG7 2RD, UK}
\author{Igor Lesanovsky}
\affiliation{Institut f\"ur Theoretische Physik, Universit\"at T\"ubingen, Auf der Morgenstelle 14, 72076 T\"ubingen, Germany}
\affiliation{School of Physics and Astronomy and \\Centre for the Mathematics and Theoretical Physics of Quantum Non-Equilibrium Systems, University of Nottingham, Nottingham, NG7 2RD, UK}

\date{\today}

\begin{abstract}
Quantum many-body systems out of equilibrium can host intriguing phenomena such as transitions to exotic dynamical states. Although this emergent behaviour can be observed in experiments, its potential for technological applications is largely unexplored. Here, we investigate the impact of collective  effects on quantum engines that extract mechanical work from a many-body system. Using an opto-mechanical cavity setup with an interacting atomic gas as a working fluid, we demonstrate theoretically that such engines produce work under periodic driving. The stationary cycle of the working fluid features nonequilibrium phase transitions, resulting in abrupt changes of the work output. Remarkably, we find that our many-body quantum engine operates even without periodic driving. This phenomenon occurs when its working fluid enters a phase that breaks continuous time-translation symmetry: the emergent time-crystalline phase can sustain the motion of a load generating mechanical work. Our findings pave the way for designing novel nonequilibrium quantum machines.
\end{abstract}

\maketitle

Future-generation nanomachines will require powerful small-scale engines whose energy output can be channeled into mechanical work storages. Proof-of-principle experiments have shown how such microscopic flywheels can be realized for working systems with few internal degrees of freedom like a single atom \cite{Thierschmann:2015aa,Rosnagel:2016aa,Josefsson:2018aa,PhysRevLett.123.080602,PhysRevLett.123.240601}. Yet it is less clear how the output of a quantum engine can be converted into motive power if its working fluid consists of a many-body system.

During the last years, much progress has been made in the design of quantum engines that operate far from equilibrium and use non-thermal sources of energy \cite{PhysRevLett.118.260603,PhysRevLett.120.260601,Niedenzu_2018,Pezzutto_2019,PhysRevB.99.024203,Abari_2019,PhysRevLett.124.170602}. The natural next step is to explore how mechanical work can be generated in such non-equilibrium settings, how collective effects, like phase transitions, affect the work output and whether they could enable novel modes of operation.

\begin{figure}[t]
\centering
\includegraphics[scale=0.61]{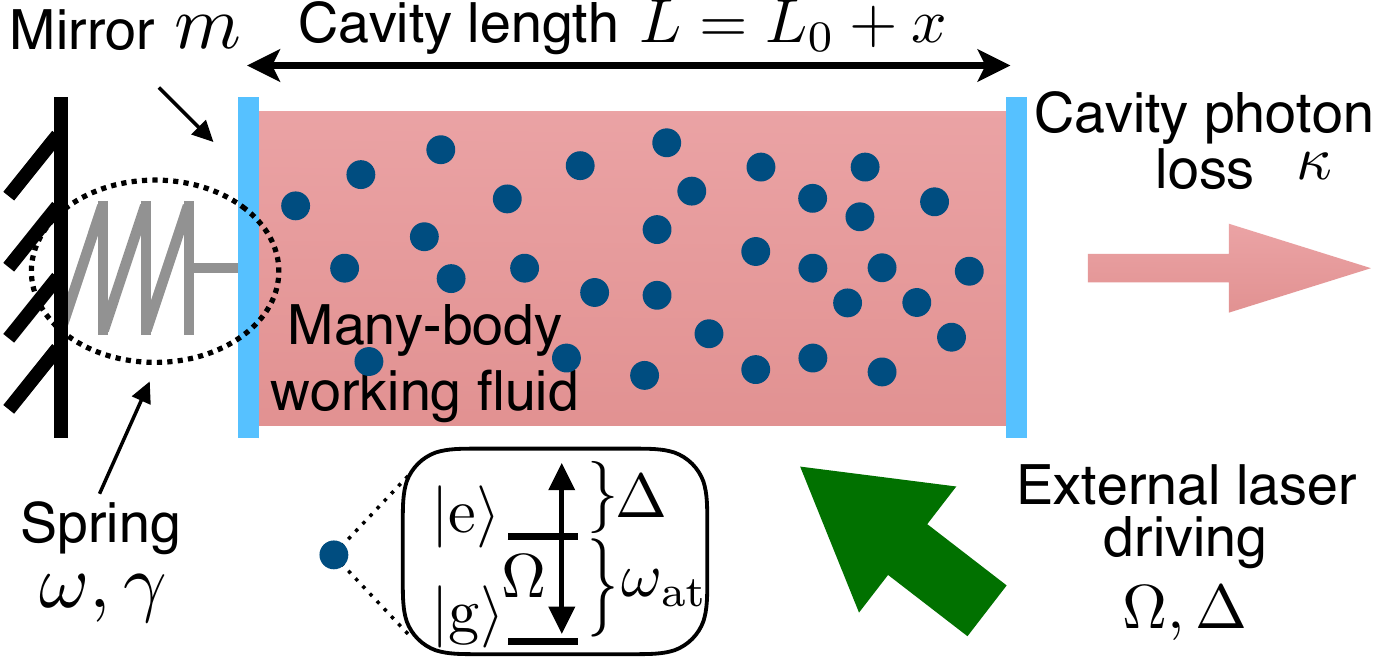}
\caption{{\bf Cavity-atom quantum engine.} The atomic working fluid is held in an optical cavity with one movable mirror (mass $m$) attached to a spring (characteristic frequency $\omega$). The cavity length $L=L_0+x$ decomposes into an equilibrium length $L_0$ and a small deviation $x$. The motion of the mirror is damped by mechanical friction (proportional to the coefficient $\gamma$) and driven by thermal fluctuations and the radiation pressure inside the cavity. Each atom is described as a two-level system with ground state $|g\rangle$, excited state $|e\rangle$ and energy splitting $\omega_{\rm at}$. Excitations are generated and destroyed through interactions with a light mode in the cavity and with the driving laser (Rabi frequency $\Omega$ and detuning $\Delta$). The cavity loss rate is $\kappa$. }
\label{Fig1}
\end{figure} 

In this article, we propose a new type of many-body quantum engine, that is driven by time-translation symmetry breaking \cite{PhysRevLett.109.160401,PhysRevLett.109.160402,PhysRevLett.121.035301} and does not require a periodic protocol. Our engine autonomously delivers mechanical work to an external load as a result of its working fluid hosting a phase with broken (continuous) time-translation symmetry, a so-called time crystal  \cite{PhysRevLett.109.160401,PhysRevLett.109.160402,PhysRevLett.121.035301}. We show that such an exotic device can be implemented with a general cavity-atom setting, which can, in principle, be realized in experiments with cold atoms \cite{Norciae1601231,Norcia259,RevModPhys.85.553}, see the sketch in Fig.\ref{Fig1}. In this setup, one mirror of the cavity is fixed, while the other one is attached to a micro-spring and can move around its equilibrium position \cite{RevModPhys.86.1391,PhysRevLett.112.150602,Brunelli_2015}. By driving the atoms inside the cavity with a periodically modulated laser, the free mirror can be forced into sustained oscillations from which we determine the work delivered by the engine \cite{Sekimoto:1997aa,10.1143/PTPS.130.17,sekimoto2010stochastic,Seifert_2012}. 

Our numerical analysis reveals two quite remarkable effects.  First, the system features a series of nonequilibrium phase transitions leading to sudden changes of the asymptotic cycle. Second, even in the absence of an explicit periodic driving, where one would \emph{a priori} expect the system to approach to a stationary state, the oscillatory motion of the mirror can be sustained as the atomic working fluid forms a time-crystalline phase \cite{PhysRevLett.109.160401,PhysRevLett.109.160402,PhysRevLett.121.035301} for properly chosen parameters.

Beyond illuminating these intriguing many-body effects, our approach has the key advantage that it admits a clear thermodynamic interpretation. Since the mirror is effectively classical, its position can be monitored without disturbing the operation cycle of the engine, thus avoiding the subtleties of measuring a quantum working fluid. Such quasi-classical work measurements make it possible to unambiguously determine the effective output of a quantum engine. At the same time, they open new ways to probe collective phenomena in nonequilibrium quantum many-body systems. \\

{ \bf \em Cavity-atom setup.--} We consider the setup of Fig.~\ref{Fig1}. An ensemble of $N$ atoms is loaded into an optical cavity with one movable mirror. Each atom is described as a two-level system with ground state $\ket{\rm g}$, excited state $\ket{\rm e}$, and level splitting $\omega_{\rm at}$. A single light mode is resonant with the cavity at frequency $\omega_{\rm cav}$. The exchange of photons between atoms and light field is described by the coupling Hamiltonian 
\begin{equation}
H_{\rm int}=\hbar \frac{g}{\sqrt{N}}\left(a\, S_++a^\dagger S_-\right) \quad \mbox{ with }\quad S_\pm=\sum_{k=1}^{N}\sigma_\pm^{(k)}\, .
\label{Ham_int}
\end{equation}
Here, $a$ and $a^\dagger$ are the photon creation and annihilation operators and $\sigma_-=\ket{\rm g} \bra{\rm e}$ and $\sigma_+=\sigma_-^\dagger$ are the atomic transition operators. The interaction strength is rescaled by the factor $1/\sqrt{N}$ as is common for light-matter interactions of this type \cite{PhysRevA.8.2517,HEPP1973360,Kirton2019}. The atoms are further driven by an external laser, whose frequency is shifted from $\omega_{\rm at}$ by the detuning $\Delta$. In the rotating frame of the laser, the atomic Hamiltonian is given by \cite{RevModPhys.85.553,PhysRevLett.118.133604,Sawant:2017aa,PhysRevLett.108.023602}
\begin{equation}
H_{\rm L}=\hbar \left[\Omega \left(S_{+}+S_{-}\right)-\frac{\Delta}{2} S_{ z}\right]\quad \mbox{ with }\quad S_{z}=\sum_{k=1}^{N}\sigma_{z}^{(k)}\, , 
\label{las-driving}
\end{equation}
where $\sigma_{ z}=\ket{\rm e}\bra{\rm e}-\ket{\rm g}\bra{\rm g}$. The Rabi frequency $\Omega$ is determined by the strength of the coherent driving. In the same rotating frame, the free Hamiltonian of the light field reads $H_{\rm ph}=-\hbar \delta\,  a^\dagger a$, where $\delta=\omega_{\rm at}+\Delta-\omega_{\rm cav}$ is the effective detuning of the cavity mode. The loss of photons from the cavity, at rate $\kappa$, is described by the dissipation super operator \cite{lindblad1976,gorini1976bis,breuer02a,Gardiner2004}
$$
\mathcal{D}_{\rm ph}[\rho]=\hbar \kappa \left(a\rho \, a^\dagger -\frac{1}{2}\left\{\rho,a^\dagger a \right\}\right)\, .
$$
In the Schr\"odinger picture, the bare photon Hamiltonian is given by $H_{\rm ph}^{\rm S}=\hbar \omega_{\rm cav}a^\dagger a$. The frequency of the photons is connected to the length $L$ of the cavity through the resonance condition $\omega_{\rm cav}=nc/(2L)$ with $n$ being an integer and $c$ the speed of light. We decompose the length of the cavity, $L=L_0+x$, into an equilibrium  contribution $L_0$ and a deviation $x$, which accounts for small oscillations of the first mirror, see Fig.~\ref{Fig1}. Expanding the Hamiltonian $H_{\rm ph}^{\rm S}$ to first order in $x/L_0$ yields  \cite{RevModPhys.86.1391,PhysRevLett.112.150602,Brunelli_2015}
\begin{equation}
H_{\rm ph}^{\rm S}\approx\hbar \omega_{\rm cav}^0\left(1-\frac{x}{L_0}\right)a^\dagger a \quad \mbox{with}\quad \omega_{\rm cav}^0=\frac{nc}{2L_0}\, .
\label{Bare-ph-energy}
\end{equation}
This result shows that the position of the mirror and the number of  photons are coupled \cite{RevModPhys.86.1391,PhysRevLett.112.150602,Brunelli_2015}. In fact, the Hamiltonian in Eq.~\eqref{Bare-ph-energy} describes a mechanical force on the mirror, which emerges from the radiation pressure inside the cavity. 

In addition, the light field in the cavity mediates an effective excitation-exchange coupling between the atoms, which arises from the interaction Hamiltonian \eqref{Ham_int} when the electromagnetic field is traced out \cite{PhysRevA.56.2249,PhysRevLett.107.277201}. In the weak-coupling regime $\kappa\gg g/\sqrt{N}$, the state of the atoms $\rho_t$ follows an effective Lindblad equation \cite{Norcia259,lindblad1976,gorini1976bis,breuer02a,Gardiner2004,SM} 
\begin{equation}
 \dot{\rho}_t=-\frac{i}{\hbar}[\tilde{H},\rho_t]+\frac{1}{\hbar}\tilde{\mathcal{D}}[\rho_t]\, .
\label{QME}
\end{equation}
Upon neglecting second-order contributions in the relative displacement $x/L_0$, the corresponding effective Hamiltonian and dissipation super-operator become
\begin{equation}
\tilde {H}=H_{\rm L}+\frac{\hbar g}{N}\left(C_0-\frac{C_1}{L_0}x\right)\, S_+S_-\, , 
\label{eff-H}
\end{equation}
and 
\begin{equation}
\tilde{\mathcal{D}}[\rho]=\frac{\hbar g}{N}\left(\Gamma_0-\frac{\Gamma_1}{L_0}x\right) \left(S_-\rho \, S_+ -\frac{1}{2}\left\{\rho,S_+S_-\right\}\right)\, ,
\label{eff-D}
\end{equation}
with the dimensionless constants \cite{SM}
\begin{align*}
&C_0=\frac{4\delta_0g}{\kappa^2+4\delta_0^2}\, ,&\Gamma_0=\frac{4\kappa g }{\kappa^2 +4\delta_0^2}\, \\
&C_1=\frac{\omega_{\rm cav}^0}{\delta_0}C_0\frac{4\delta_0^2-\kappa^2}{4\delta_0^2+\kappa^2}\, , &\Gamma_1=\omega_{\rm cav}^0\Gamma_0\frac{8\delta_0}{\kappa^2 +4\delta_0^2}
\end{align*}
and the detuning parameter $\delta_0=\omega_{\rm at}+\Delta-\omega_{\rm cav}^0$. \\

{ \bf \em Dynamics of the mirror.--} The mirror is a massive object, whose ground state energy is small compared to the typical energy of thermal excitations. That is, we have $\hbar\omega\ll k_{{{\rm B}}} T$, where $\omega$ is the characteristic frequency of the spring attached to the mirror, see Fig.~\ref{Fig1}, and $T$ is the base temperature of the setup. The position of the mirror can thus be treated as a classical degree of freedom, whose dynamics is governed by the Langevin equation \cite{Seifert_2012,Yaghoubi_2017}
\begin{equation}
m\ddot{x}+\gamma \dot{x}+m\omega^2 \,x=f_t+\xi_t\, ;
\label{UD-sto-eq}
\end{equation}
here, $\gamma$ is the damping coefficient, $m$ is the mass of the mirror and the stochastic force $\xi_t$, which describes thermal fluctuations, obeys $\mathbb{E}[\xi_t]=0$ and $\mathbb{E}[\xi_t\xi_{s}]=2\gamma k_{\rm B}T\delta\left(t-s\right)$. The symbol $\mathbb{E}$ indicates average over all realizations \cite{risken1996fokker}.

The deterministic force $f_t=-\langle \left[\frac{\partial}{\partial x}H_{\rm ph}^{\rm S} \right]\rangle_t$ is due to the light-mediated coupling between the mirror and the working fluid. In the effective picture of an interacting atomic gas, it is, up to second-order corrections in the displacement of the mirror $x/L_0$, given by \cite{SM},  
\begin{equation}
f_t\approx \hbar \frac{g}{N}\frac{\omega_{\rm cav}^0}{\kappa \, L_0}\left(\Gamma_0-\Gamma_1\frac{x}{L_0}\right) \braket{S_+S_-}_t\, ,
\label{sys-for}
\end{equation}
where angular brackets denote the average with respect to the state of the atomic system $\rho_t$. Together with this relation, the effective master equation \eqref{QME} and the Langevin equation \eqref{UD-sto-eq} determine the joint dynamics of the mirror and the working fluid. \\

\begin{figure}[t]
\centering
\includegraphics[scale=0.27]{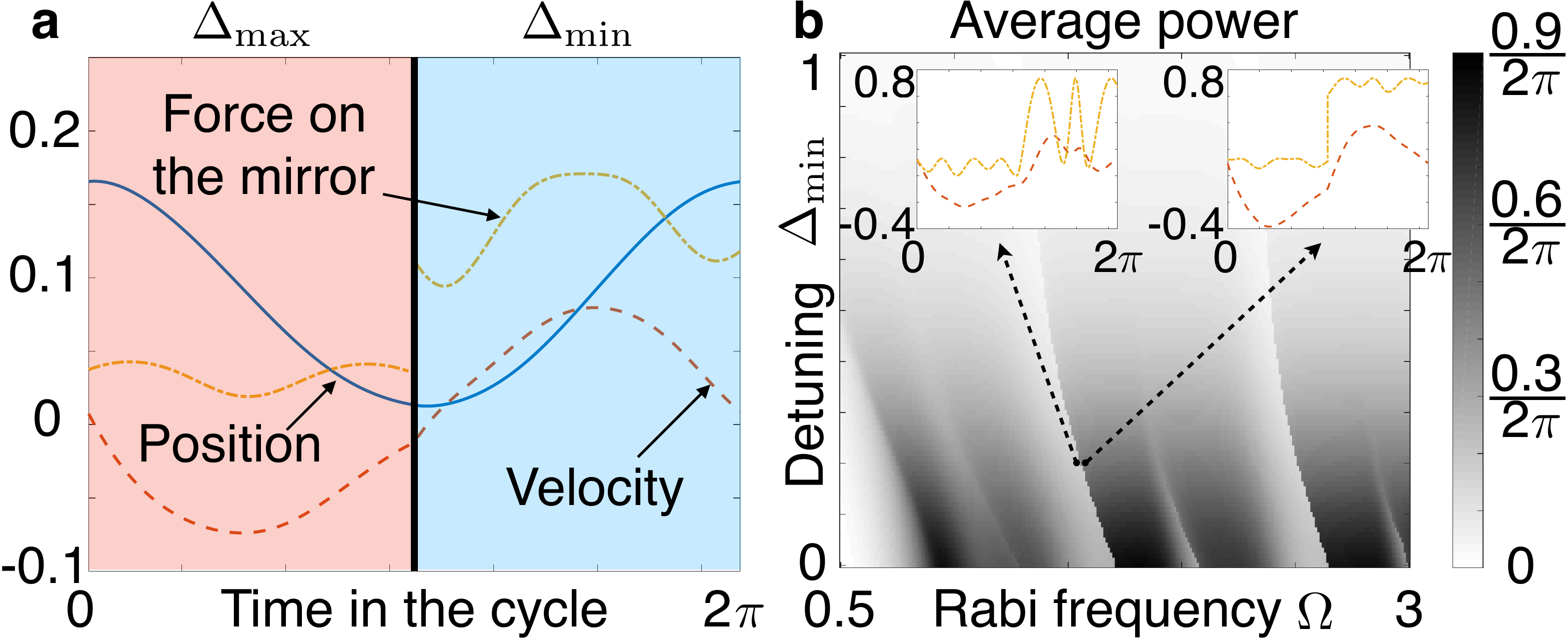}
\caption{{\bf Periodic driving.} (a) Periodic motion of the mirror driven by the many-body engine. The average velocity of the mirror, $\bar{v}_t$, is given in units of ${\rm v}_0=\frac{\omega_{\rm cav}^0}{\omega}\frac{\hbar D_0}{m }$, the average position, $\bar{x}_t$, in units of ${\rm v}_0/\omega$ and the force in units of $\hbar \omega_{\rm cav}^0 D_0$. Time is given in units of $\omega^{-1}$. 
This representative cycle is obtained for $\omega_{\rm at}-\omega_{\rm cav}^0=0.1\omega$,  $\Delta_{\rm max}=2\omega$ and $\Delta_{\rm min}=\kappa=g=\omega$. The mean velocity is non-zero, proving that the engine constantly delivers energy to the mirror through the force $f_t$. (b) Power output, in units of $m{\rm v}_0^2\omega$, as a function of $\Omega/\omega$ and $\Delta_{\rm min}/\omega$ with fixed $\Delta_{\rm max}-\Delta_{\rm min}=\omega$. In the insets, we show the mean velocity and the force for $\Delta_{\rm min}=0.2\omega$ and two slightly different values of the Rabi frequency, $\Omega/\omega=1.55,1.56$, for which the power output differs substantially. The force profile changes abruptly from an oscillatory pattern to a two plateau-like shape indicating a non-equilibrium phase transition. Numerical results have been obtained by simulating the dynamics of the mirror for sufficiently long times, such that the system has converged to its asymptotic cycle. }
\label{Fig2}
\end{figure} 

{ \bf \em Finite-density limit.--} We now consider the limit of large atom numbers, $N\gg1$, focussing on the case where the linear density of atoms in the cavity, $D_0=N/L_0$, is fixed. This assumption, which is typically well justified in experiments, makes it possible to simplify our mathematical model. First, the constants $C_1$ and $\Gamma_1$, appearing in Eqs.~\eqref{eff-H}-\eqref{eff-D}, become irrelevant for the dynamics, as they are of order $N^{-2}$, and can thus be neglected. Second, the normalized correlation functions $\langle S_+S_-\rangle/N^2$ factorize, since emergent correlations between different atoms are wiped out in the large-$N$ limit \cite{BENATTI2016381,Benatti_2018}. That is, we have $\langle S_+ S_+\rangle/N^2\sim s_+s_-$ with $s_\pm = \lim_{N\to\infty}\langle S_\pm\rangle/N$. 

As a result, the collective atomic variables $s_\pm$ and $s_z=\lim_{N\to\infty}\langle S_z\rangle/N$ obey the mean-field type dynamical equations \cite{Benatti_2018,SM}
\begin{equation}
\begin{split}
\dot{s}_+&=-i\Omega \, s_{ z}-i\Delta s_+-igC_0\, s_zs_++\frac{g_0\, \Gamma_0}{2}s_{ z}s_+\, ,\\
\dot{s}_{ z}&=2i\Omega\left(s_--s_+\right)-2g\Gamma_0s_+s_-\, ,
\end{split}
\label{mf-eqs}
\end{equation}
and the Langevin equation \eqref{UD-sto-eq} becomes 
\begin{equation}
m\ddot{x}+\gamma \dot{x}+m\omega^2 \,x=\hbar \frac{g\omega_{\rm cav}^0\Gamma_0D_0}{\kappa }s_+s_-+\xi_t\, ;
\label{UD-sto-eq2}
\end{equation}
since the expression \eqref{sys-for} for the deterministic force reduces to 
\begin{equation}
f_t=\hbar \frac{g\omega_{\rm cav}^0\Gamma_0 D_0}{\kappa }s_+s_-
\label{force-fin-den}
\end{equation}
in the thermodynamic limit. The Eqs.\eqref{mf-eqs}-\eqref{UD-sto-eq2} provide a complete dynamical model of our engine in terms of the four variables $s_\pm$, $s_z$ and $x$. \\

{ \bf \em Work extraction through the mirror.--} We first consider a conventional isothermal engine cycle, where the detuning $\Delta$ is periodically modulated to provide energy input. For simplicity, we focus on a quench protocol, where $\Delta=\Delta_{\rm max}$ during the first half of the period and $\Delta=\Delta_{\rm min}$ during the second half, as shown in Fig.~\ref{Fig2}(a). The period $t_{\rm c}$ of the driving is resonant with the eigenfrequency of the mirror, $t_{\rm c}=2\pi/\omega$. As a consequence of the driving, the state of the atoms $\rho_t$ and the average position of the mirror approach an asymptotic cycle with period $t_{\rm c}$. As initial conditions we consider all atoms in their ground state and the mirror in its equilibrium position ($x_0=\dot{x}_0=0$). However, at least at a qualitative level, our results  do not depend on the specific initial conditions.  

Owing to energy conservation, the average amount of work that is transferred from the atomic system to the mirror plus the energy contribution of the thermal fluctuations must be equal to the average heat dissipation due to mechanical friction. Hence, the power delivered by the engine per cycle is given by  
\begin{equation}
P_{\rm av}=\frac{\gamma}{t_{\rm c}}\int_0^{t_{\rm c}}dt  \left(\mathbb{E}\left[v_t^2\right]-\frac{k_{\rm B} T}{m}\right)=\frac{\gamma}{t_{\rm c}}\int_0^{t_{\rm c}}dt \, \bar{v}_t^2\, ,
\label{P-cyc}
\end{equation}
as shown in Methods by using tools of stochastic thermodynamics. Here, $v_t=\dot{x}_t$ is the velocity of the mirror, $\gamma k_{\rm B}T/m$ represents the thermal energy dissipated by mechanical friction, and $\bar{v}_t$ is the average velocity of the mirror --which can be obtained from Eq.~\eqref{UD-sto-eq2} with $\xi_t=0$. 

In Fig.~\ref{Fig2}(b), the generated mean power $P_{\rm av}$ is plotted as a function of the Rabi frequency $\Omega$ and the lower level of the detuning $\Delta_{{{\rm min}}}$. We find that $P_{\rm av}$ is positive over a large range of parameters. This result proves that our engine is able to produce usable work by sustaining the periodic motion of the mirror against constant damping. Quite remarkably, the average power output features discontinuous jumps signalling nonequilibrium phase transitions in the asymptotic periodic state as illustrated by the insets of Fig.~\ref{Fig2}(b). This new type of phase transition generalizes steady-state nonequilibrium ones to periodically driven settings. The power output acts as an order parameter which can be used to unveil the occurence of sudden changes in the asymptotic periodic dynamics of the many-body working fluid.\\

\begin{figure}[t]
\centering
\includegraphics[scale=0.27]{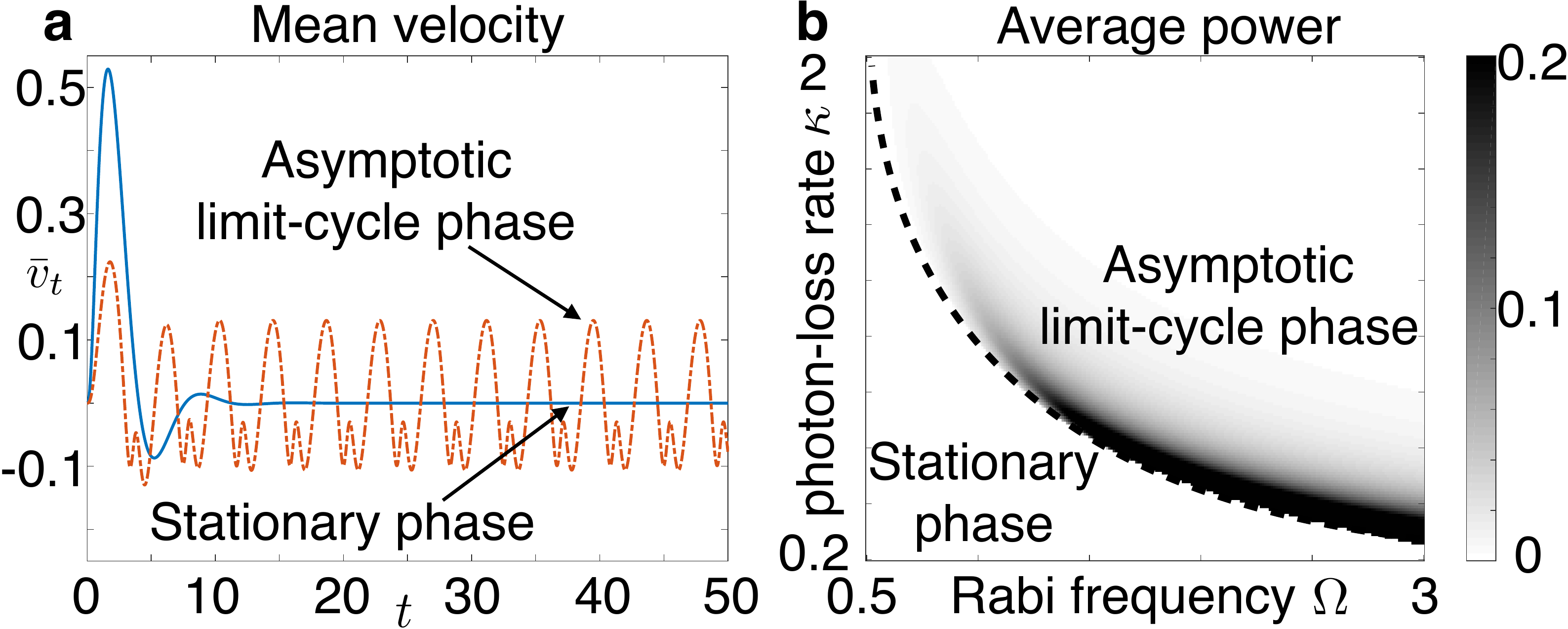}
\caption{{\bf Time-crystal quantum engine. } (a) Mean velocity of the mirror in units of ${\rm v}_0=\frac{\omega_{\rm cav}^0}{\omega}\frac{\hbar D_0}{m }$ as a function of time (in units of $\omega^{-1}$) for two different photon-loss rates. For $\kappa/\omega =0.5$, the mirror comes to rest while sustained oscillations emerge for $\kappa/\omega=1.5$. For this plot we have set $\Delta=0$, $\omega_{\rm at}-\omega_{\rm cav}^0=0.1\omega$ and $\Omega=g=\omega$. (b) Average power output, in units of $m{\rm v}_0^2\omega$, as a function of the photon-loss rate $\kappa$ and the Rabi frequency $\Omega$, both in units of $\omega$. Along the dashed line $\kappa\Omega\sim 1$, a phase transition occurs, where the average power jumps to a finite value as the working fluid spontaneously forms a time crystal. The scale has been truncated at $0.2$, but significantly larger values for the work output ($P_{\rm av}>2$) are found. The maximum value, for the chosen parameters, is given by $P_{\rm av}\sim2.8$.}
\label{Fig3}
\end{figure}

{ \bf \em Time-translation symmetry breaking.--} Once the time-dependent modulation of the detuning is turned off, one would expect the mirror to come to rest as the working fluid settles to a steady state. However, our  analysis shows that, for properly chosen parameters, the engine still drives sustained oscillations of the mirror, even if the detuning is fixed. This \emph{a priori} surprising phenomenon arises as a consequence of the working fluid entering a time-crystal phase, which breaks the continuous temporal translation symmetry of the time-independent generator of the dynamics \cite{PhysRevLett.121.035301}. The engine thereby acquires a new operation mechanism, which does not require cyclic control protocols and instead makes it possible to generate periodic motion from steady-state driving, as illustrated in Fig.~\ref{Fig3}(a).

In the absence of a periodic protocol, there is no natural recurrence time for the long-time dynamics which, in general, may or may not approach an asymptotic cycle. To explore this regime quantitatively, we thus need to determine the average power of the engine by calculating the average heat loss generated by the mirror over a long time window. Namely, we compute the power as 
$$
P_{\rm av}=\lim_{t_{\rm obs}\to \infty}\frac{\gamma}{t_{\rm obs}}\int_0^{t_{\rm obs}} dt \, \bar{v}_t^2\, .
$$
If the position of the mirror settles on an asymptotic cycle with a well-defined period, this definition coincides with the one given in Eq.~\eqref{P-cyc}.

The results of our analysis are summarized in Fig.~\ref{Fig3}(b). In the weak photon-loss regime, i.e.~for $\kappa\ll \Omega$, the working fluid settles to a stationary state where no mechanical work is produced. Approximately at $\kappa \sim 1/\Omega$, for the specific choice of parameters, a phase transition occurs and the average power abruptly increases as the mirror breaks into sustained oscillations. This effect is most pronounced at moderate photon-loss rates, where our time-crystal engine delivers the largest output. 

The working mechanism of this new operation mode of the engine can be understood as follows. The effective dissipation constant $\Gamma_0$ decays with large photon-loss rates $\kappa$. In this regime, characterized by weak dissipation on the atoms ($\Gamma_0\ll1$), the dynamics of the working fluid is dominated by the Rabi driving at frequency $\Omega$. As a result, the steady-state manifold of the atomic system becomes degenerate and long-lived oscillations within this manifold emerge \cite{PhysRevLett.121.035301}. Thus, oscillating coherences are responsible for the time-dependent force on the mirror. When atomic dissipation dominates, coherent oscillations are suppressed and the working fluid approaches a time-invariant steady-state. In this case, the radiation pressure on the mirror is constant and the mirror comes to rest. \\

{ \bf \em Discussion.--} We have developed a general framework for the dynamical description of many-body quantum engines, which includes the external load as a semi-classical degree of freedom. This approach makes it possible to determine the performance of the engine directly by monitoring the coupled dynamics of both the working fluid and of the load. 

\noindent This perspective allowed us to obtain two key results. First, in the periodic mode of operation, where the engine is driven by modulations of an external control parameter, a new type of nonequilibrium phase transitions emerges. The power output of the engine thereby plays the role of an order parameter. 

\noindent Second, we have demonstrated that, even when the engine is not driven through a periodic protocol, and is thus described by a time-independent generator, it can still deliver mechanical work. The emergence of this new regime is due to a nonequilibrium phase transition in the atomic working fluid towards an exotic state that features sustained coherent oscillations. This dynamical time-crystalline phase \cite{PhysRevLett.109.160401,PhysRevLett.109.160402,PhysRevLett.121.035301} can drive the motion of the load without relying on a time-dependent control protocol. Our approach paves the way to explore new mechanisms of power generation enabled by collective many-body effects and, at the same time, provides a natural description of many-body quantum engines. 

\noindent Our predictions on the power output of our engine can be tested with current technology in cavity-atom experiments \cite{Mekhov:2007aa,Brennecke235,Baumann:2010aa,Norciae1601231,Norcia259}, and our general numerical analysis covers a wide range of different setups. 

\begin{acknowledgments}
\noindent FC acknowledges support through a Teach@T\"ubingen Fellowship. KB received support from the University of Nottingham through a Nottingham Research Fellowship and from UK Research and Innovation through a Future Leaders Fellowship (Grant Reference: MR/S034714/1). IL acknowledges support from the DFG through SPP 1929 (GiRyd) as well as from the ``Wissenschaftler-R\"uckkehrprogramm GSO/CZS" of the Carl-Zeiss-Stiftung and the German Scholars Organization e.V. 
\end{acknowledgments}

\bibliography{Notes_BIBLIO}{}

\begin{thebibliography}{45}%
\makeatletter
\providecommand \@ifxundefined [1]{%
 \@ifx{#1\undefined}
}%
\providecommand \@ifnum [1]{%
 \ifnum #1\expandafter \@firstoftwo
 \else \expandafter \@secondoftwo
 \fi
}%
\providecommand \@ifx [1]{%
 \ifx #1\expandafter \@firstoftwo
 \else \expandafter \@secondoftwo
 \fi
}%
\providecommand \natexlab [1]{#1}%
\providecommand \enquote  [1]{``#1''}%
\providecommand \bibnamefont  [1]{#1}%
\providecommand \bibfnamefont [1]{#1}%
\providecommand \citenamefont [1]{#1}%
\providecommand \href@noop [0]{\@secondoftwo}%
\providecommand \href [0]{\begingroup \@sanitize@url \@href}%
\providecommand \@href[1]{\@@startlink{#1}\@@href}%
\providecommand \@@href[1]{\endgroup#1\@@endlink}%
\providecommand \@sanitize@url [0]{\catcode `\\12\catcode `\$12\catcode
  `\&12\catcode `\#12\catcode `\^12\catcode `\_12\catcode `\%12\relax}%
\providecommand \@@startlink[1]{}%
\providecommand \@@endlink[0]{}%
\providecommand \url  [0]{\begingroup\@sanitize@url \@url }%
\providecommand \@url [1]{\endgroup\@href {#1}{\urlprefix }}%
\providecommand \urlprefix  [0]{URL }%
\providecommand \Eprint [0]{\href }%
\providecommand \doibase [0]{http://dx.doi.org/}%
\providecommand \selectlanguage [0]{\@gobble}%
\providecommand \bibinfo  [0]{\@secondoftwo}%
\providecommand \bibfield  [0]{\@secondoftwo}%
\providecommand \translation [1]{[#1]}%
\providecommand \BibitemOpen [0]{}%
\providecommand \bibitemStop [0]{}%
\providecommand \bibitemNoStop [0]{.\EOS\space}%
\providecommand \EOS [0]{\spacefactor3000\relax}%
\providecommand \BibitemShut  [1]{\csname bibitem#1\endcsname}%
\let\auto@bib@innerbib\@empty
\bibitem [{\citenamefont {Thierschmann}\ \emph {et~al.}(2015)\citenamefont
  {Thierschmann}, \citenamefont {S{\'a}nchez}, \citenamefont {Sothmann},
  \citenamefont {Arnold}, \citenamefont {Heyn}, \citenamefont {Hansen},
  \citenamefont {Buhmann},\ and\ \citenamefont
  {Molenkamp}}]{Thierschmann:2015aa}%
  \BibitemOpen
  \bibfield  {author} {\bibinfo {author} {\bibfnamefont {Holger}\ \bibnamefont
  {Thierschmann}}, \bibinfo {author} {\bibfnamefont {Rafael}\ \bibnamefont
  {S{\'a}nchez}}, \bibinfo {author} {\bibfnamefont {Bj{\"o}rn}\ \bibnamefont
  {Sothmann}}, \bibinfo {author} {\bibfnamefont {Fabian}\ \bibnamefont
  {Arnold}}, \bibinfo {author} {\bibfnamefont {Christian}\ \bibnamefont
  {Heyn}}, \bibinfo {author} {\bibfnamefont {Wolfgang}\ \bibnamefont {Hansen}},
  \bibinfo {author} {\bibfnamefont {Hartmut}\ \bibnamefont {Buhmann}}, \ and\
  \bibinfo {author} {\bibfnamefont {Laurens~W.}\ \bibnamefont {Molenkamp}},\
  }\bibfield  {title} {\enquote {\bibinfo {title} {Three-terminal energy
  harvester with coupled quantum dots},}\ }\href {\doibase
  10.1038/nnano.2015.176} {\bibfield  {journal} {\bibinfo  {journal} {Nature
  Nanotechnology}\ }\textbf {\bibinfo {volume} {10}},\ \bibinfo {pages}
  {854--858} (\bibinfo {year} {2015})}\BibitemShut {NoStop}%
\bibitem [{\citenamefont {Ro{\ss}nagel}\ \emph {et~al.}(2016)\citenamefont
  {Ro{\ss}nagel}, \citenamefont {Dawkins}, \citenamefont {Tolazzi},
  \citenamefont {Abah}, \citenamefont {Lutz}, \citenamefont {Schmidt-Kaler},\
  and\ \citenamefont {Singer}}]{Rosnagel:2016aa}%
  \BibitemOpen
  \bibfield  {author} {\bibinfo {author} {\bibfnamefont {Johannes}\
  \bibnamefont {Ro{\ss}nagel}}, \bibinfo {author} {\bibfnamefont {Samuel~T.}\
  \bibnamefont {Dawkins}}, \bibinfo {author} {\bibfnamefont {Karl~N.}\
  \bibnamefont {Tolazzi}}, \bibinfo {author} {\bibfnamefont {Obinna}\
  \bibnamefont {Abah}}, \bibinfo {author} {\bibfnamefont {Eric}\ \bibnamefont
  {Lutz}}, \bibinfo {author} {\bibfnamefont {Ferdinand}\ \bibnamefont
  {Schmidt-Kaler}}, \ and\ \bibinfo {author} {\bibfnamefont {Kilian}\
  \bibnamefont {Singer}},\ }\bibfield  {title} {\enquote {\bibinfo {title} {A
  single-atom heat engine},}\ }\href {\doibase 10.1126/science.aad6320}
  {\bibfield  {journal} {\bibinfo  {journal} {Science}\ }\textbf {\bibinfo
  {volume} {352}},\ \bibinfo {pages} {325} (\bibinfo {year}
  {2016})}\BibitemShut {NoStop}%
\bibitem [{\citenamefont {Josefsson}\ \emph {et~al.}(2018)\citenamefont
  {Josefsson}, \citenamefont {Svilans}, \citenamefont {Burke}, \citenamefont
  {Hoffmann}, \citenamefont {Fahlvik}, \citenamefont {Thelander}, \citenamefont
  {Leijnse},\ and\ \citenamefont {Linke}}]{Josefsson:2018aa}%
  \BibitemOpen
  \bibfield  {author} {\bibinfo {author} {\bibfnamefont {Martin}\ \bibnamefont
  {Josefsson}}, \bibinfo {author} {\bibfnamefont {Artis}\ \bibnamefont
  {Svilans}}, \bibinfo {author} {\bibfnamefont {Adam~M.}\ \bibnamefont
  {Burke}}, \bibinfo {author} {\bibfnamefont {Eric~A.}\ \bibnamefont
  {Hoffmann}}, \bibinfo {author} {\bibfnamefont {Sofia}\ \bibnamefont
  {Fahlvik}}, \bibinfo {author} {\bibfnamefont {Claes}\ \bibnamefont
  {Thelander}}, \bibinfo {author} {\bibfnamefont {Martin}\ \bibnamefont
  {Leijnse}}, \ and\ \bibinfo {author} {\bibfnamefont {Heiner}\ \bibnamefont
  {Linke}},\ }\bibfield  {title} {\enquote {\bibinfo {title} {A quantum-dot
  heat engine operating close to the thermodynamic efficiency limits},}\ }\href
  {\doibase 10.1038/s41565-018-0200-5} {\bibfield  {journal} {\bibinfo
  {journal} {Nature Nanotechnology}\ }\textbf {\bibinfo {volume} {13}},\
  \bibinfo {pages} {920--924} (\bibinfo {year} {2018})}\BibitemShut {NoStop}%
\bibitem [{\citenamefont {von Lindenfels}\ \emph {et~al.}(2019)\citenamefont
  {von Lindenfels}, \citenamefont {Gr\"ab}, \citenamefont {Schmiegelow},
  \citenamefont {Kaushal}, \citenamefont {Schulz}, \citenamefont {Mitchison},
  \citenamefont {Goold}, \citenamefont {Schmidt-Kaler},\ and\ \citenamefont
  {Poschinger}}]{PhysRevLett.123.080602}%
  \BibitemOpen
  \bibfield  {author} {\bibinfo {author} {\bibfnamefont {D.}~\bibnamefont {von
  Lindenfels}}, \bibinfo {author} {\bibfnamefont {O.}~\bibnamefont {Gr\"ab}},
  \bibinfo {author} {\bibfnamefont {C.~T.}\ \bibnamefont {Schmiegelow}},
  \bibinfo {author} {\bibfnamefont {V.}~\bibnamefont {Kaushal}}, \bibinfo
  {author} {\bibfnamefont {J.}~\bibnamefont {Schulz}}, \bibinfo {author}
  {\bibfnamefont {Mark~T.}\ \bibnamefont {Mitchison}}, \bibinfo {author}
  {\bibfnamefont {John}\ \bibnamefont {Goold}}, \bibinfo {author}
  {\bibfnamefont {F.}~\bibnamefont {Schmidt-Kaler}}, \ and\ \bibinfo {author}
  {\bibfnamefont {U.~G.}\ \bibnamefont {Poschinger}},\ }\bibfield  {title}
  {\enquote {\bibinfo {title} {Spin heat engine coupled to a
  harmonic-oscillator flywheel},}\ }\href {\doibase
  10.1103/PhysRevLett.123.080602} {\bibfield  {journal} {\bibinfo  {journal}
  {Phys. Rev. Lett.}\ }\textbf {\bibinfo {volume} {123}},\ \bibinfo {pages}
  {080602} (\bibinfo {year} {2019})}\BibitemShut {NoStop}%
\bibitem [{\citenamefont {Peterson}\ \emph {et~al.}(2019)\citenamefont
  {Peterson}, \citenamefont {Batalh\~ao}, \citenamefont {Herrera},
  \citenamefont {Souza}, \citenamefont {Sarthour}, \citenamefont {Oliveira},\
  and\ \citenamefont {Serra}}]{PhysRevLett.123.240601}%
  \BibitemOpen
  \bibfield  {author} {\bibinfo {author} {\bibfnamefont {John P.~S.}\
  \bibnamefont {Peterson}}, \bibinfo {author} {\bibfnamefont {Tiago~B.}\
  \bibnamefont {Batalh\~ao}}, \bibinfo {author} {\bibfnamefont {Marcela}\
  \bibnamefont {Herrera}}, \bibinfo {author} {\bibfnamefont {Alexandre~M.}\
  \bibnamefont {Souza}}, \bibinfo {author} {\bibfnamefont {Roberto~S.}\
  \bibnamefont {Sarthour}}, \bibinfo {author} {\bibfnamefont {Ivan~S.}\
  \bibnamefont {Oliveira}}, \ and\ \bibinfo {author} {\bibfnamefont
  {Roberto~M.}\ \bibnamefont {Serra}},\ }\bibfield  {title} {\enquote {\bibinfo
  {title} {Experimental characterization of a spin quantum heat engine},}\
  }\href {\doibase 10.1103/PhysRevLett.123.240601} {\bibfield  {journal}
  {\bibinfo  {journal} {Phys. Rev. Lett.}\ }\textbf {\bibinfo {volume} {123}},\
  \bibinfo {pages} {240601} (\bibinfo {year} {2019})}\BibitemShut {NoStop}%
\bibitem [{\citenamefont {Elouard}\ \emph {et~al.}(2017)\citenamefont
  {Elouard}, \citenamefont {Herrera-Mart\'{\i}}, \citenamefont {Huard},\ and\
  \citenamefont {Auff\`eves}}]{PhysRevLett.118.260603}%
  \BibitemOpen
  \bibfield  {author} {\bibinfo {author} {\bibfnamefont {Cyril}\ \bibnamefont
  {Elouard}}, \bibinfo {author} {\bibfnamefont {David}\ \bibnamefont
  {Herrera-Mart\'{\i}}}, \bibinfo {author} {\bibfnamefont {Benjamin}\
  \bibnamefont {Huard}}, \ and\ \bibinfo {author} {\bibfnamefont {Alexia}\
  \bibnamefont {Auff\`eves}},\ }\bibfield  {title} {\enquote {\bibinfo {title}
  {Extracting work from quantum measurement in {Maxwell's Demon} engines},}\
  }\href {\doibase 10.1103/PhysRevLett.118.260603} {\bibfield  {journal}
  {\bibinfo  {journal} {Phys. Rev. Lett.}\ }\textbf {\bibinfo {volume} {118}},\
  \bibinfo {pages} {260603} (\bibinfo {year} {2017})}\BibitemShut {NoStop}%
\bibitem [{\citenamefont {Elouard}\ and\ \citenamefont
  {Jordan}(2018)}]{PhysRevLett.120.260601}%
  \BibitemOpen
  \bibfield  {author} {\bibinfo {author} {\bibfnamefont {Cyril}\ \bibnamefont
  {Elouard}}\ and\ \bibinfo {author} {\bibfnamefont {Andrew~N.}\ \bibnamefont
  {Jordan}},\ }\bibfield  {title} {\enquote {\bibinfo {title} {Efficient
  quantum measurement engines},}\ }\href {\doibase
  10.1103/PhysRevLett.120.260601} {\bibfield  {journal} {\bibinfo  {journal}
  {Phys. Rev. Lett.}\ }\textbf {\bibinfo {volume} {120}},\ \bibinfo {pages}
  {260601} (\bibinfo {year} {2018})}\BibitemShut {NoStop}%
\bibitem [{\citenamefont {Niedenzu}\ and\ \citenamefont
  {Kurizki}(2018)}]{Niedenzu_2018}%
  \BibitemOpen
  \bibfield  {author} {\bibinfo {author} {\bibfnamefont {Wolfgang}\
  \bibnamefont {Niedenzu}}\ and\ \bibinfo {author} {\bibfnamefont {Gershon}\
  \bibnamefont {Kurizki}},\ }\bibfield  {title} {\enquote {\bibinfo {title}
  {Cooperative many-body enhancement of quantum thermal machine power},}\
  }\href {\doibase 10.1088/1367-2630/aaed55} {\bibfield  {journal} {\bibinfo
  {journal} {New Journal of Physics}\ }\textbf {\bibinfo {volume} {20}},\
  \bibinfo {pages} {113038} (\bibinfo {year} {2018})}\BibitemShut {NoStop}%
\bibitem [{\citenamefont {Pezzutto}\ \emph {et~al.}(2019)\citenamefont
  {Pezzutto}, \citenamefont {Paternostro},\ and\ \citenamefont
  {Omar}}]{Pezzutto_2019}%
  \BibitemOpen
  \bibfield  {author} {\bibinfo {author} {\bibfnamefont {Marco}\ \bibnamefont
  {Pezzutto}}, \bibinfo {author} {\bibfnamefont {Mauro}\ \bibnamefont
  {Paternostro}}, \ and\ \bibinfo {author} {\bibfnamefont {Yasser}\
  \bibnamefont {Omar}},\ }\bibfield  {title} {\enquote {\bibinfo {title} {An
  out-of-equilibrium {non-Markovian} quantum heat engine},}\ }\href {\doibase
  10.1088/2058-9565/aaf5b4} {\bibfield  {journal} {\bibinfo  {journal} {Quantum
  Science and Technology}\ }\textbf {\bibinfo {volume} {4}},\ \bibinfo {pages}
  {025002} (\bibinfo {year} {2019})}\BibitemShut {NoStop}%
\bibitem [{\citenamefont {Yunger~Halpern}\ \emph {et~al.}(2019)\citenamefont
  {Yunger~Halpern}, \citenamefont {White}, \citenamefont {Gopalakrishnan},\
  and\ \citenamefont {Refael}}]{PhysRevB.99.024203}%
  \BibitemOpen
  \bibfield  {author} {\bibinfo {author} {\bibfnamefont {Nicole}\ \bibnamefont
  {Yunger~Halpern}}, \bibinfo {author} {\bibfnamefont {Christopher~David}\
  \bibnamefont {White}}, \bibinfo {author} {\bibfnamefont {Sarang}\
  \bibnamefont {Gopalakrishnan}}, \ and\ \bibinfo {author} {\bibfnamefont
  {Gil}\ \bibnamefont {Refael}},\ }\bibfield  {title} {\enquote {\bibinfo
  {title} {Quantum engine based on many-body localization},}\ }\href {\doibase
  10.1103/PhysRevB.99.024203} {\bibfield  {journal} {\bibinfo  {journal} {Phys.
  Rev. B}\ }\textbf {\bibinfo {volume} {99}},\ \bibinfo {pages} {024203}
  (\bibinfo {year} {2019})}\BibitemShut {NoStop}%
\bibitem [{\citenamefont {Abari}\ \emph {et~al.}(2019)\citenamefont {Abari},
  \citenamefont {Angelis}, \citenamefont {Zippilli},\ and\ \citenamefont
  {Vitali}}]{Abari_2019}%
  \BibitemOpen
  \bibfield  {author} {\bibinfo {author} {\bibfnamefont {Najmeh~Etehadi}\
  \bibnamefont {Abari}}, \bibinfo {author} {\bibfnamefont {Giulia Vittoria~De}\
  \bibnamefont {Angelis}}, \bibinfo {author} {\bibfnamefont {Stefano}\
  \bibnamefont {Zippilli}}, \ and\ \bibinfo {author} {\bibfnamefont {David}\
  \bibnamefont {Vitali}},\ }\bibfield  {title} {\enquote {\bibinfo {title} {An
  optomechanical heat engine with feedback-controlled in-loop light},}\ }\href
  {\doibase 10.1088/1367-2630/ab41e7} {\bibfield  {journal} {\bibinfo
  {journal} {New Journal of Physics}\ }\textbf {\bibinfo {volume} {21}},\
  \bibinfo {pages} {093051} (\bibinfo {year} {2019})}\BibitemShut {NoStop}%
\bibitem [{\citenamefont {Carollo}\ \emph {et~al.}(2020)\citenamefont
  {Carollo}, \citenamefont {Gambetta}, \citenamefont {Brandner}, \citenamefont
  {Garrahan},\ and\ \citenamefont {Lesanovsky}}]{PhysRevLett.124.170602}%
  \BibitemOpen
  \bibfield  {author} {\bibinfo {author} {\bibfnamefont {Federico}\
  \bibnamefont {Carollo}}, \bibinfo {author} {\bibfnamefont {Filippo~M.}\
  \bibnamefont {Gambetta}}, \bibinfo {author} {\bibfnamefont {Kay}\
  \bibnamefont {Brandner}}, \bibinfo {author} {\bibfnamefont {Juan~P.}\
  \bibnamefont {Garrahan}}, \ and\ \bibinfo {author} {\bibfnamefont {Igor}\
  \bibnamefont {Lesanovsky}},\ }\bibfield  {title} {\enquote {\bibinfo {title}
  {Nonequilibrium quantum many-body {Rydberg} atom engine},}\ }\href {\doibase
  10.1103/PhysRevLett.124.170602} {\bibfield  {journal} {\bibinfo  {journal}
  {Phys. Rev. Lett.}\ }\textbf {\bibinfo {volume} {124}},\ \bibinfo {pages}
  {170602} (\bibinfo {year} {2020})}\BibitemShut {NoStop}%
\bibitem [{\citenamefont {Wilczek}(2012)}]{PhysRevLett.109.160401}%
  \BibitemOpen
  \bibfield  {author} {\bibinfo {author} {\bibfnamefont {Frank}\ \bibnamefont
  {Wilczek}},\ }\bibfield  {title} {\enquote {\bibinfo {title} {Quantum time
  crystals},}\ }\href {\doibase 10.1103/PhysRevLett.109.160401} {\bibfield
  {journal} {\bibinfo  {journal} {Phys. Rev. Lett.}\ }\textbf {\bibinfo
  {volume} {109}},\ \bibinfo {pages} {160401} (\bibinfo {year}
  {2012})}\BibitemShut {NoStop}%
\bibitem [{\citenamefont {Shapere}\ and\ \citenamefont
  {Wilczek}(2012)}]{PhysRevLett.109.160402}%
  \BibitemOpen
  \bibfield  {author} {\bibinfo {author} {\bibfnamefont {Alfred}\ \bibnamefont
  {Shapere}}\ and\ \bibinfo {author} {\bibfnamefont {Frank}\ \bibnamefont
  {Wilczek}},\ }\bibfield  {title} {\enquote {\bibinfo {title} {Classical time
  crystals},}\ }\href {\doibase 10.1103/PhysRevLett.109.160402} {\bibfield
  {journal} {\bibinfo  {journal} {Phys. Rev. Lett.}\ }\textbf {\bibinfo
  {volume} {109}},\ \bibinfo {pages} {160402} (\bibinfo {year}
  {2012})}\BibitemShut {NoStop}%
\bibitem [{\citenamefont {Iemini}\ \emph {et~al.}(2018)\citenamefont {Iemini},
  \citenamefont {Russomanno}, \citenamefont {Keeling}, \citenamefont
  {Schir\`o}, \citenamefont {Dalmonte},\ and\ \citenamefont
  {Fazio}}]{PhysRevLett.121.035301}%
  \BibitemOpen
  \bibfield  {author} {\bibinfo {author} {\bibfnamefont {F.}~\bibnamefont
  {Iemini}}, \bibinfo {author} {\bibfnamefont {A.}~\bibnamefont {Russomanno}},
  \bibinfo {author} {\bibfnamefont {J.}~\bibnamefont {Keeling}}, \bibinfo
  {author} {\bibfnamefont {M.}~\bibnamefont {Schir\`o}}, \bibinfo {author}
  {\bibfnamefont {M.}~\bibnamefont {Dalmonte}}, \ and\ \bibinfo {author}
  {\bibfnamefont {R.}~\bibnamefont {Fazio}},\ }\bibfield  {title} {\enquote
  {\bibinfo {title} {Boundary time crystals},}\ }\href {\doibase
  10.1103/PhysRevLett.121.035301} {\bibfield  {journal} {\bibinfo  {journal}
  {Phys. Rev. Lett.}\ }\textbf {\bibinfo {volume} {121}},\ \bibinfo {pages}
  {035301} (\bibinfo {year} {2018})}\BibitemShut {NoStop}%
\bibitem [{\citenamefont {Norcia}\ \emph {et~al.}(2016)\citenamefont {Norcia},
  \citenamefont {Winchester}, \citenamefont {Cline},\ and\ \citenamefont
  {Thompson}}]{Norciae1601231}%
  \BibitemOpen
  \bibfield  {author} {\bibinfo {author} {\bibfnamefont {Matthew~A.}\
  \bibnamefont {Norcia}}, \bibinfo {author} {\bibfnamefont {Matthew~N.}\
  \bibnamefont {Winchester}}, \bibinfo {author} {\bibfnamefont {Julia R.~K.}\
  \bibnamefont {Cline}}, \ and\ \bibinfo {author} {\bibfnamefont {James~K.}\
  \bibnamefont {Thompson}},\ }\bibfield  {title} {\enquote {\bibinfo {title}
  {Superradiance on the millihertz linewidth strontium clock transition},}\
  }\href {\doibase 10.1126/sciadv.1601231} {\bibfield  {journal} {\bibinfo
  {journal} {Science Advances}\ }\textbf {\bibinfo {volume} {2}} (\bibinfo
  {year} {2016}),\ 10.1126/sciadv.1601231}\BibitemShut {NoStop}%
\bibitem [{\citenamefont {Norcia}\ \emph {et~al.}(2018)\citenamefont {Norcia},
  \citenamefont {Lewis-Swan}, \citenamefont {Cline}, \citenamefont {Zhu},
  \citenamefont {Rey},\ and\ \citenamefont {Thompson}}]{Norcia259}%
  \BibitemOpen
  \bibfield  {author} {\bibinfo {author} {\bibfnamefont {Matthew~A.}\
  \bibnamefont {Norcia}}, \bibinfo {author} {\bibfnamefont {Robert~J.}\
  \bibnamefont {Lewis-Swan}}, \bibinfo {author} {\bibfnamefont {Julia R.~K.}\
  \bibnamefont {Cline}}, \bibinfo {author} {\bibfnamefont {Bihui}\ \bibnamefont
  {Zhu}}, \bibinfo {author} {\bibfnamefont {Ana~M.}\ \bibnamefont {Rey}}, \
  and\ \bibinfo {author} {\bibfnamefont {James~K.}\ \bibnamefont {Thompson}},\
  }\bibfield  {title} {\enquote {\bibinfo {title} {Cavity-mediated collective
  spin-exchange interactions in a strontium superradiant laser},}\ }\href
  {\doibase 10.1126/science.aar3102} {\bibfield  {journal} {\bibinfo  {journal}
  {Science}\ }\textbf {\bibinfo {volume} {361}},\ \bibinfo {pages} {259--262}
  (\bibinfo {year} {2018})}\BibitemShut {NoStop}%
\bibitem [{\citenamefont {Ritsch}\ \emph {et~al.}(2013)\citenamefont {Ritsch},
  \citenamefont {Domokos}, \citenamefont {Brennecke},\ and\ \citenamefont
  {Esslinger}}]{RevModPhys.85.553}%
  \BibitemOpen
  \bibfield  {author} {\bibinfo {author} {\bibfnamefont {Helmut}\ \bibnamefont
  {Ritsch}}, \bibinfo {author} {\bibfnamefont {Peter}\ \bibnamefont {Domokos}},
  \bibinfo {author} {\bibfnamefont {Ferdinand}\ \bibnamefont {Brennecke}}, \
  and\ \bibinfo {author} {\bibfnamefont {Tilman}\ \bibnamefont {Esslinger}},\
  }\bibfield  {title} {\enquote {\bibinfo {title} {Cold atoms in
  cavity-generated dynamical optical potentials},}\ }\href {\doibase
  10.1103/RevModPhys.85.553} {\bibfield  {journal} {\bibinfo  {journal} {Rev.
  Mod. Phys.}\ }\textbf {\bibinfo {volume} {85}},\ \bibinfo {pages} {553--601}
  (\bibinfo {year} {2013})}\BibitemShut {NoStop}%
\bibitem [{\citenamefont {Aspelmeyer}\ \emph {et~al.}(2014)\citenamefont
  {Aspelmeyer}, \citenamefont {Kippenberg},\ and\ \citenamefont
  {Marquardt}}]{RevModPhys.86.1391}%
  \BibitemOpen
  \bibfield  {author} {\bibinfo {author} {\bibfnamefont {Markus}\ \bibnamefont
  {Aspelmeyer}}, \bibinfo {author} {\bibfnamefont {Tobias~J.}\ \bibnamefont
  {Kippenberg}}, \ and\ \bibinfo {author} {\bibfnamefont {Florian}\
  \bibnamefont {Marquardt}},\ }\bibfield  {title} {\enquote {\bibinfo {title}
  {Cavity optomechanics},}\ }\href {\doibase 10.1103/RevModPhys.86.1391}
  {\bibfield  {journal} {\bibinfo  {journal} {Rev. Mod. Phys.}\ }\textbf
  {\bibinfo {volume} {86}},\ \bibinfo {pages} {1391--1452} (\bibinfo {year}
  {2014})}\BibitemShut {NoStop}%
\bibitem [{\citenamefont {Zhang}\ \emph {et~al.}(2014)\citenamefont {Zhang},
  \citenamefont {Bariani},\ and\ \citenamefont
  {Meystre}}]{PhysRevLett.112.150602}%
  \BibitemOpen
  \bibfield  {author} {\bibinfo {author} {\bibfnamefont {Keye}\ \bibnamefont
  {Zhang}}, \bibinfo {author} {\bibfnamefont {Francesco}\ \bibnamefont
  {Bariani}}, \ and\ \bibinfo {author} {\bibfnamefont {Pierre}\ \bibnamefont
  {Meystre}},\ }\bibfield  {title} {\enquote {\bibinfo {title} {Quantum
  optomechanical heat engine},}\ }\href {\doibase
  10.1103/PhysRevLett.112.150602} {\bibfield  {journal} {\bibinfo  {journal}
  {Phys. Rev. Lett.}\ }\textbf {\bibinfo {volume} {112}},\ \bibinfo {pages}
  {150602} (\bibinfo {year} {2014})}\BibitemShut {NoStop}%
\bibitem [{\citenamefont {Brunelli}\ \emph {et~al.}(2015)\citenamefont
  {Brunelli}, \citenamefont {Xuereb}, \citenamefont {Ferraro}, \citenamefont
  {Chiara}, \citenamefont {Kiesel},\ and\ \citenamefont
  {Paternostro}}]{Brunelli_2015}%
  \BibitemOpen
  \bibfield  {author} {\bibinfo {author} {\bibfnamefont {M}~\bibnamefont
  {Brunelli}}, \bibinfo {author} {\bibfnamefont {A}~\bibnamefont {Xuereb}},
  \bibinfo {author} {\bibfnamefont {A}~\bibnamefont {Ferraro}}, \bibinfo
  {author} {\bibfnamefont {G~De}\ \bibnamefont {Chiara}}, \bibinfo {author}
  {\bibfnamefont {N}~\bibnamefont {Kiesel}}, \ and\ \bibinfo {author}
  {\bibfnamefont {M}~\bibnamefont {Paternostro}},\ }\bibfield  {title}
  {\enquote {\bibinfo {title} {Out-of-equilibrium thermodynamics of quantum
  optomechanical systems},}\ }\href {\doibase 10.1088/1367-2630/17/3/035016}
  {\bibfield  {journal} {\bibinfo  {journal} {New Journal of Physics}\ }\textbf
  {\bibinfo {volume} {17}},\ \bibinfo {pages} {035016} (\bibinfo {year}
  {2015})}\BibitemShut {NoStop}%
\bibitem [{\citenamefont {Sekimoto}(1997)}]{Sekimoto:1997aa}%
  \BibitemOpen
  \bibfield  {author} {\bibinfo {author} {\bibfnamefont {Ken}\ \bibnamefont
  {Sekimoto}},\ }\bibfield  {title} {\enquote {\bibinfo {title} {Kinetic
  characterization of heat bath and the energetics of thermal ratchet
  models},}\ }\bibfield  {booktitle} {\emph {\bibinfo {booktitle} {Journal of
  the Physical Society of Japan}},\ }\href {\doibase 10.1143/JPSJ.66.1234}
  {\bibfield  {journal} {\bibinfo  {journal} {Journal of the Physical Society
  of Japan}\ }\textbf {\bibinfo {volume} {66}},\ \bibinfo {pages} {1234--1237}
  (\bibinfo {year} {1997})}\BibitemShut {NoStop}%
\bibitem [{\citenamefont {Sekimoto}(1998)}]{10.1143/PTPS.130.17}%
  \BibitemOpen
  \bibfield  {author} {\bibinfo {author} {\bibfnamefont {Ken}\ \bibnamefont
  {Sekimoto}},\ }\bibfield  {title} {\enquote {\bibinfo {title} {{Langevin
  Equation and Thermodynamics}},}\ }\href {\doibase 10.1143/PTPS.130.17}
  {\bibfield  {journal} {\bibinfo  {journal} {Progress of Theoretical Physics
  Supplement}\ }\textbf {\bibinfo {volume} {130}},\ \bibinfo {pages} {17--27}
  (\bibinfo {year} {1998})}\BibitemShut {NoStop}%
\bibitem [{\citenamefont {Sekimoto}(2010)}]{sekimoto2010stochastic}%
  \BibitemOpen
  \bibfield  {author} {\bibinfo {author} {\bibfnamefont {Ken}\ \bibnamefont
  {Sekimoto}},\ }\href@noop {} {\emph {\bibinfo {title} {Stochastic
  energetics}}},\ Vol.\ \bibinfo {volume} {799}\ (\bibinfo  {publisher}
  {Springer, Berlin},\ \bibinfo {year} {2010})\BibitemShut {NoStop}%
\bibitem [{\citenamefont {Seifert}(2012)}]{Seifert_2012}%
  \BibitemOpen
  \bibfield  {author} {\bibinfo {author} {\bibfnamefont {Udo}\ \bibnamefont
  {Seifert}},\ }\bibfield  {title} {\enquote {\bibinfo {title} {Stochastic
  thermodynamics, fluctuation theorems and molecular machines},}\ }\href
  {\doibase 10.1088/0034-4885/75/12/126001} {\bibfield  {journal} {\bibinfo
  {journal} {Reports on Progress in Physics}\ }\textbf {\bibinfo {volume}
  {75}},\ \bibinfo {pages} {126001} (\bibinfo {year} {2012})}\BibitemShut
  {NoStop}%
\bibitem [{\citenamefont {Hepp}\ and\ \citenamefont
  {Lieb}(1973{\natexlab{a}})}]{PhysRevA.8.2517}%
  \BibitemOpen
  \bibfield  {author} {\bibinfo {author} {\bibfnamefont {Klaus}\ \bibnamefont
  {Hepp}}\ and\ \bibinfo {author} {\bibfnamefont {Elliott~H.}\ \bibnamefont
  {Lieb}},\ }\bibfield  {title} {\enquote {\bibinfo {title} {Equilibrium
  statistical mechanics of matter interacting with the quantized radiation
  field},}\ }\href {\doibase 10.1103/PhysRevA.8.2517} {\bibfield  {journal}
  {\bibinfo  {journal} {Phys. Rev. A}\ }\textbf {\bibinfo {volume} {8}},\
  \bibinfo {pages} {2517--2525} (\bibinfo {year}
  {1973}{\natexlab{a}})}\BibitemShut {NoStop}%
\bibitem [{\citenamefont {Hepp}\ and\ \citenamefont
  {Lieb}(1973{\natexlab{b}})}]{HEPP1973360}%
  \BibitemOpen
  \bibfield  {author} {\bibinfo {author} {\bibfnamefont {Klaus}\ \bibnamefont
  {Hepp}}\ and\ \bibinfo {author} {\bibfnamefont {Elliott~H}\ \bibnamefont
  {Lieb}},\ }\bibfield  {title} {\enquote {\bibinfo {title} {On the
  superradiant phase transition for molecules in a quantized radiation field:
  the {Dicke} maser model},}\ }\href {\doibase
  https://doi.org/10.1016/0003-4916(73)90039-0} {\bibfield  {journal} {\bibinfo
   {journal} {Annals of Physics}\ }\textbf {\bibinfo {volume} {76}},\ \bibinfo
  {pages} {360 -- 404} (\bibinfo {year} {1973}{\natexlab{b}})}\BibitemShut
  {NoStop}%
\bibitem [{\citenamefont {Kirton}\ \emph {et~al.}(2019)\citenamefont {Kirton},
  \citenamefont {Roses}, \citenamefont {Keeling},\ and\ \citenamefont
  {Dalla~Torre}}]{Kirton2019}%
  \BibitemOpen
  \bibfield  {author} {\bibinfo {author} {\bibfnamefont {Peter}\ \bibnamefont
  {Kirton}}, \bibinfo {author} {\bibfnamefont {Mor~M.}\ \bibnamefont {Roses}},
  \bibinfo {author} {\bibfnamefont {Jonathan}\ \bibnamefont {Keeling}}, \ and\
  \bibinfo {author} {\bibfnamefont {Emanuele~G.}\ \bibnamefont {Dalla~Torre}},\
  }\bibfield  {title} {\enquote {\bibinfo {title} {Introduction to the {Dicke}
  model: From equilibrium to nonequilibrium, and vice versa},}\ }\href
  {\doibase 10.1002/qute.201800043} {\bibfield  {journal} {\bibinfo  {journal}
  {Advanced Quantum Technologies}\ }\textbf {\bibinfo {volume} {2}},\ \bibinfo
  {pages} {1800043} (\bibinfo {year} {2019})}\BibitemShut {NoStop}%
\bibitem [{\citenamefont {Hamsen}\ \emph {et~al.}(2017)\citenamefont {Hamsen},
  \citenamefont {Tolazzi}, \citenamefont {Wilk},\ and\ \citenamefont
  {Rempe}}]{PhysRevLett.118.133604}%
  \BibitemOpen
  \bibfield  {author} {\bibinfo {author} {\bibfnamefont {Christoph}\
  \bibnamefont {Hamsen}}, \bibinfo {author} {\bibfnamefont {Karl~Nicolas}\
  \bibnamefont {Tolazzi}}, \bibinfo {author} {\bibfnamefont {Tatjana}\
  \bibnamefont {Wilk}}, \ and\ \bibinfo {author} {\bibfnamefont {Gerhard}\
  \bibnamefont {Rempe}},\ }\bibfield  {title} {\enquote {\bibinfo {title}
  {Two-photon blockade in an atom-driven cavity {QED} system},}\ }\href
  {\doibase 10.1103/PhysRevLett.118.133604} {\bibfield  {journal} {\bibinfo
  {journal} {Phys. Rev. Lett.}\ }\textbf {\bibinfo {volume} {118}},\ \bibinfo
  {pages} {133604} (\bibinfo {year} {2017})}\BibitemShut {NoStop}%
\bibitem [{\citenamefont {Sawant}\ and\ \citenamefont
  {Rangwala}(2017)}]{Sawant:2017aa}%
  \BibitemOpen
  \bibfield  {author} {\bibinfo {author} {\bibfnamefont {Rahul}\ \bibnamefont
  {Sawant}}\ and\ \bibinfo {author} {\bibfnamefont {S.~A.}\ \bibnamefont
  {Rangwala}},\ }\bibfield  {title} {\enquote {\bibinfo {title} {Lasing by
  driven atoms-cavity system in collective strong coupling regime},}\ }\href
  {\doibase 10.1038/s41598-017-11799-5} {\bibfield  {journal} {\bibinfo
  {journal} {Scientific Reports}\ }\textbf {\bibinfo {volume} {7}},\ \bibinfo
  {pages} {11432} (\bibinfo {year} {2017})}\BibitemShut {NoStop}%
\bibitem [{\citenamefont {Lee}\ \emph {et~al.}(2012)\citenamefont {Lee},
  \citenamefont {H\"affner},\ and\ \citenamefont
  {Cross}}]{PhysRevLett.108.023602}%
  \BibitemOpen
  \bibfield  {author} {\bibinfo {author} {\bibfnamefont {Tony~E.}\ \bibnamefont
  {Lee}}, \bibinfo {author} {\bibfnamefont {H.}~\bibnamefont {H\"affner}}, \
  and\ \bibinfo {author} {\bibfnamefont {M.~C.}\ \bibnamefont {Cross}},\
  }\bibfield  {title} {\enquote {\bibinfo {title} {Collective quantum jumps of
  {Rydberg} atoms},}\ }\href {\doibase 10.1103/PhysRevLett.108.023602}
  {\bibfield  {journal} {\bibinfo  {journal} {Phys. Rev. Lett.}\ }\textbf
  {\bibinfo {volume} {108}},\ \bibinfo {pages} {023602} (\bibinfo {year}
  {2012})}\BibitemShut {NoStop}%
\bibitem [{\citenamefont {Lindblad}(1976)}]{lindblad1976}%
  \BibitemOpen
  \bibfield  {author} {\bibinfo {author} {\bibfnamefont {G.}~\bibnamefont
  {Lindblad}},\ }\bibfield  {title} {\enquote {\bibinfo {title} {On the
  generators of quantum dynamical semigroups},}\ }\href
  {https://projecteuclid.org:443/euclid.cmp/1103899849} {\bibfield  {journal}
  {\bibinfo  {journal} {Comm. Math. Phys.}\ }\textbf {\bibinfo {volume} {48}},\
  \bibinfo {pages} {119--130} (\bibinfo {year} {1976})}\BibitemShut {NoStop}%
\bibitem [{\citenamefont {Gorini}\ \emph {et~al.}(1976)\citenamefont {Gorini},
  \citenamefont {Kossakowski},\ and\ \citenamefont
  {Sudarshan}}]{gorini1976bis}%
  \BibitemOpen
  \bibfield  {author} {\bibinfo {author} {\bibfnamefont {Vittorio}\
  \bibnamefont {Gorini}}, \bibinfo {author} {\bibfnamefont {Andrzej}\
  \bibnamefont {Kossakowski}}, \ and\ \bibinfo {author} {\bibfnamefont
  {Ennackal Chandy~George}\ \bibnamefont {Sudarshan}},\ }\bibfield  {title}
  {\enquote {\bibinfo {title} {Completely positive dynamical semigroups of
  {N-level} systems},}\ }\href {\doibase 10.1063/1.522979} {\bibfield
  {journal} {\bibinfo  {journal} {Journal of Mathematical Physics}\ }\textbf
  {\bibinfo {volume} {17}},\ \bibinfo {pages} {821--825} (\bibinfo {year}
  {1976})}\BibitemShut {NoStop}%
\bibitem [{\citenamefont {Breuer}\ and\ \citenamefont
  {Petruccione}(2002)}]{breuer02a}%
  \BibitemOpen
  \bibfield  {author} {\bibinfo {author} {\bibfnamefont {H.P.}\ \bibnamefont
  {Breuer}}\ and\ \bibinfo {author} {\bibfnamefont {F.}~\bibnamefont
  {Petruccione}},\ }\href@noop {} {\emph {\bibinfo {title} {The theory of open
  quantum systems}}}\ (\bibinfo  {publisher} {Oxford University Press},\
  \bibinfo {year} {2002})\BibitemShut {NoStop}%
\bibitem [{\citenamefont {Gardiner}\ and\ \citenamefont
  {Zoller}(2004)}]{Gardiner2004}%
  \BibitemOpen
  \bibfield  {author} {\bibinfo {author} {\bibfnamefont {Crispin}\ \bibnamefont
  {Gardiner}}\ and\ \bibinfo {author} {\bibfnamefont {Peter}\ \bibnamefont
  {Zoller}},\ }\href@noop {} {\emph {\bibinfo {title} {{Quantum noise}}}}\
  (\bibinfo  {publisher} {Springer},\ \bibinfo {year} {2004})\BibitemShut
  {NoStop}%
\bibitem [{\citenamefont {Agarwal}\ \emph {et~al.}(1997)\citenamefont
  {Agarwal}, \citenamefont {Puri},\ and\ \citenamefont
  {Singh}}]{PhysRevA.56.2249}%
  \BibitemOpen
  \bibfield  {author} {\bibinfo {author} {\bibfnamefont {G.~S.}\ \bibnamefont
  {Agarwal}}, \bibinfo {author} {\bibfnamefont {R.~R.}\ \bibnamefont {Puri}}, \
  and\ \bibinfo {author} {\bibfnamefont {R.~P.}\ \bibnamefont {Singh}},\
  }\bibfield  {title} {\enquote {\bibinfo {title} {Atomic schr\"odinger cat
  states},}\ }\href {\doibase 10.1103/PhysRevA.56.2249} {\bibfield  {journal}
  {\bibinfo  {journal} {Phys. Rev. A}\ }\textbf {\bibinfo {volume} {56}},\
  \bibinfo {pages} {2249--2254} (\bibinfo {year} {1997})}\BibitemShut {NoStop}%
\bibitem [{\citenamefont {Gopalakrishnan}\ \emph {et~al.}(2011)\citenamefont
  {Gopalakrishnan}, \citenamefont {Lev},\ and\ \citenamefont
  {Goldbart}}]{PhysRevLett.107.277201}%
  \BibitemOpen
  \bibfield  {author} {\bibinfo {author} {\bibfnamefont {Sarang}\ \bibnamefont
  {Gopalakrishnan}}, \bibinfo {author} {\bibfnamefont {Benjamin~L.}\
  \bibnamefont {Lev}}, \ and\ \bibinfo {author} {\bibfnamefont {Paul~M.}\
  \bibnamefont {Goldbart}},\ }\bibfield  {title} {\enquote {\bibinfo {title}
  {Frustration and glassiness in spin models with cavity-mediated
  interactions},}\ }\href {\doibase 10.1103/PhysRevLett.107.277201} {\bibfield
  {journal} {\bibinfo  {journal} {Phys. Rev. Lett.}\ }\textbf {\bibinfo
  {volume} {107}},\ \bibinfo {pages} {277201} (\bibinfo {year}
  {2011})}\BibitemShut {NoStop}%
\bibitem [{SM()}]{SM}%
  \BibitemOpen
  \href@noop {} {\bibinfo  {journal} {See supplemental material for details.}\
  }\BibitemShut {NoStop}%
\bibitem [{\citenamefont {Yaghoubi}\ \emph {et~al.}(2017)\citenamefont
  {Yaghoubi}, \citenamefont {Foulaadvand}, \citenamefont {B{\'{e}}rut},\ and\
  \citenamefont {{\L}uczka}}]{Yaghoubi_2017}%
  \BibitemOpen
\bibfield  {journal} {  }\bibfield  {author} {\bibinfo {author} {\bibfnamefont
  {Mohammad}\ \bibnamefont {Yaghoubi}}, \bibinfo {author} {\bibfnamefont
  {M~Ebrahim}\ \bibnamefont {Foulaadvand}}, \bibinfo {author} {\bibfnamefont
  {Antoine}\ \bibnamefont {B{\'{e}}rut}}, \ and\ \bibinfo {author}
  {\bibfnamefont {Jerzy}\ \bibnamefont {{\L}uczka}},\ }\bibfield  {title}
  {\enquote {\bibinfo {title} {Energetics of a driven brownian harmonic
  oscillator},}\ }\href {\doibase 10.1088/1742-5468/aa9346} {\bibfield
  {journal} {\bibinfo  {journal} {Journal of Statistical Mechanics: Theory and
  Experiment}\ }\textbf {\bibinfo {volume} {2017}},\ \bibinfo {pages} {113206}
  (\bibinfo {year} {2017})}\BibitemShut {NoStop}%
\bibitem [{\citenamefont {Risken}(1996)}]{risken1996fokker}%
  \BibitemOpen
  \bibfield  {author} {\bibinfo {author} {\bibfnamefont {Hannes}\ \bibnamefont
  {Risken}},\ }\href@noop {} {\emph {\bibinfo {title} {The Fokker-Planck
  Equation}}}\ (\bibinfo  {publisher} {Springer, Berlin},\ \bibinfo {year}
  {1996})\BibitemShut {NoStop}%
\bibitem [{\citenamefont {Benatti}\ \emph {et~al.}(2016)\citenamefont
  {Benatti}, \citenamefont {Carollo}, \citenamefont {Floreanini},\ and\
  \citenamefont {Narnhofer}}]{BENATTI2016381}%
  \BibitemOpen
  \bibfield  {author} {\bibinfo {author} {\bibfnamefont {F.}~\bibnamefont
  {Benatti}}, \bibinfo {author} {\bibfnamefont {F.}~\bibnamefont {Carollo}},
  \bibinfo {author} {\bibfnamefont {R.}~\bibnamefont {Floreanini}}, \ and\
  \bibinfo {author} {\bibfnamefont {H.}~\bibnamefont {Narnhofer}},\ }\bibfield
  {title} {\enquote {\bibinfo {title} {Non-{Markovian} mesoscopic dissipative
  dynamics of open quantum spin chains},}\ }\href {\doibase
  https://doi.org/10.1016/j.physleta.2015.10.062} {\bibfield  {journal}
  {\bibinfo  {journal} {Physics Letters A}\ }\textbf {\bibinfo {volume}
  {380}},\ \bibinfo {pages} {381 -- 389} (\bibinfo {year} {2016})}\BibitemShut
  {NoStop}%
\bibitem [{\citenamefont {Benatti}\ \emph {et~al.}(2018)\citenamefont
  {Benatti}, \citenamefont {Carollo}, \citenamefont {Floreanini},\ and\
  \citenamefont {Narnhofer}}]{Benatti_2018}%
  \BibitemOpen
  \bibfield  {author} {\bibinfo {author} {\bibfnamefont {F}~\bibnamefont
  {Benatti}}, \bibinfo {author} {\bibfnamefont {F}~\bibnamefont {Carollo}},
  \bibinfo {author} {\bibfnamefont {R}~\bibnamefont {Floreanini}}, \ and\
  \bibinfo {author} {\bibfnamefont {H}~\bibnamefont {Narnhofer}},\ }\bibfield
  {title} {\enquote {\bibinfo {title} {Quantum spin chain dissipative
  mean-field dynamics},}\ }\href {\doibase 10.1088/1751-8121/aacbdb} {\bibfield
   {journal} {\bibinfo  {journal} {Journal of Physics A: Mathematical and
  Theoretical}\ }\textbf {\bibinfo {volume} {51}},\ \bibinfo {pages} {325001}
  (\bibinfo {year} {2018})}\BibitemShut {NoStop}%
\bibitem [{\citenamefont {Mekhov}\ \emph {et~al.}(2007)\citenamefont {Mekhov},
  \citenamefont {Maschler},\ and\ \citenamefont {Ritsch}}]{Mekhov:2007aa}%
  \BibitemOpen
  \bibfield  {author} {\bibinfo {author} {\bibfnamefont {Igor~B.}\ \bibnamefont
  {Mekhov}}, \bibinfo {author} {\bibfnamefont {Christoph}\ \bibnamefont
  {Maschler}}, \ and\ \bibinfo {author} {\bibfnamefont {Helmut}\ \bibnamefont
  {Ritsch}},\ }\bibfield  {title} {\enquote {\bibinfo {title} {Probing quantum
  phases of ultracold atoms in optical lattices by transmission spectra in
  cavity quantum electrodynamics},}\ }\href {\doibase 10.1038/nphys571}
  {\bibfield  {journal} {\bibinfo  {journal} {Nature Physics}\ }\textbf
  {\bibinfo {volume} {3}},\ \bibinfo {pages} {319--323} (\bibinfo {year}
  {2007})}\BibitemShut {NoStop}%
\bibitem [{\citenamefont {Brennecke}\ \emph {et~al.}(2008)\citenamefont
  {Brennecke}, \citenamefont {Ritter}, \citenamefont {Donner},\ and\
  \citenamefont {Esslinger}}]{Brennecke235}%
  \BibitemOpen
  \bibfield  {author} {\bibinfo {author} {\bibfnamefont {Ferdinand}\
  \bibnamefont {Brennecke}}, \bibinfo {author} {\bibfnamefont {Stephan}\
  \bibnamefont {Ritter}}, \bibinfo {author} {\bibfnamefont {Tobias}\
  \bibnamefont {Donner}}, \ and\ \bibinfo {author} {\bibfnamefont {Tilman}\
  \bibnamefont {Esslinger}},\ }\bibfield  {title} {\enquote {\bibinfo {title}
  {Cavity optomechanics with a {Bose-Einstein} condensate},}\ }\href {\doibase
  10.1126/science.1163218} {\bibfield  {journal} {\bibinfo  {journal}
  {Science}\ }\textbf {\bibinfo {volume} {322}},\ \bibinfo {pages} {235--238}
  (\bibinfo {year} {2008})}\BibitemShut {NoStop}%
\bibitem [{\citenamefont {Baumann}\ \emph {et~al.}(2010)\citenamefont
  {Baumann}, \citenamefont {Guerlin}, \citenamefont {Brennecke},\ and\
  \citenamefont {Esslinger}}]{Baumann:2010aa}%
  \BibitemOpen
  \bibfield  {author} {\bibinfo {author} {\bibfnamefont {Kristian}\
  \bibnamefont {Baumann}}, \bibinfo {author} {\bibfnamefont {Christine}\
  \bibnamefont {Guerlin}}, \bibinfo {author} {\bibfnamefont {Ferdinand}\
  \bibnamefont {Brennecke}}, \ and\ \bibinfo {author} {\bibfnamefont {Tilman}\
  \bibnamefont {Esslinger}},\ }\bibfield  {title} {\enquote {\bibinfo {title}
  {Dicke quantum phase transition with a superfluid gas in an optical
  cavity},}\ }\href {\doibase 10.1038/nature09009} {\bibfield  {journal}
  {\bibinfo  {journal} {Nature}\ }\textbf {\bibinfo {volume} {464}},\ \bibinfo
  {pages} {1301--1306} (\bibinfo {year} {2010})}\BibitemShut {NoStop}%
\end{thebibliography}%

\section*{Methods}
\noindent {\it \bf Stochastic dynamics of the mirror in the finite-density regime.} Owing to its large mass, the mirror can be described as a classical oscillator, which is driven by the force from the atoms and subject to thermal fluctuations due to its environment having a finite temperature $T$. The displacement $x$ of the mirror from its equilibrium position follows the underdamped Langevin equation \cite{Seifert_2012} in Eq.~\eqref{UD-sto-eq2}. The systematic force $f_t$ stems from the radiation pressure inside the cavity and is related to the state of the atoms via the expression \eqref{force-fin-den}. 

A convenient way to solve Eq.~\eqref{UD-sto-eq2}, is to rewrite it as a system of first-order differential equations, 
\begin{equation}
\frac{d}{dt}\begin{pmatrix}
x_t\\ v_t
\end{pmatrix}=\begin{pmatrix}
0&1\\
-\omega^2&-2\gamma_{0}
\end{pmatrix}\begin{pmatrix}
x_t\\ v_t
\end{pmatrix}+\frac{1}{m}\begin{pmatrix}
0\\ f_t+\xi_t
\end{pmatrix}
\label{sys-first-order}
\end{equation}
where $v_t=\dot{x}_t$ and $\gamma_0=\gamma/(2m)$. Upon introducing the vector notation 
$$
\vec{x}_t=\begin{pmatrix}
x_t\\ v_t
\end{pmatrix}\, ,\quad \vec{G}_t=\frac{1}{m}\begin{pmatrix}
0\\ f_t
\end{pmatrix}\, , \quad M=\begin{pmatrix}
0&1\\-\omega^2&-2\gamma_0
\end{pmatrix}\, ,
$$
the formal solution of Eq.~\eqref{sys-first-order} can be written as 
$$
\vec{x}_t=e^{t\, M}\vec{x}_0+\int_0^tds\, e^{(t-s)M}\vec{G}_s\, .
$$
With initial conditions $x_0=v_0=0$, we have  
$$
x_t=\bar{x}_t+x_t^{\xi}\, ,
$$
where (defining $\Sigma=\sqrt{\omega^2-\gamma_0^2}$)
\begin{equation}
\bar{x}_t=\int_0^tds \, \frac{1}{m\Sigma}e^{-\gamma_0(t-s)}\sin\left[\left(t-s\right)\Sigma\right]f_s\, ,
\label{mean-position}
\end{equation}
is the mean position of the mirror, and 
\begin{equation}
x_t^\xi=\int_0^tds \, \frac{1}{m\Sigma}e^{-\gamma_0(t-s)}\sin\left[\left(t-s\right)\Sigma\right]\xi_s\, ,
\label{fluc-position}
\end{equation}
corresponds to the fluctuating component of the position. As is explicit in the above equations, the mean position follows the driving force, while the fluctuating term depends on the thermal noise $\xi_t$ and has zero average. The analogous decomposition for the velocity of the mirror reads
$$
v_t=\bar{v}_t+v_t^{\xi}\, ,
$$
where the average velocity is given by 
\begin{equation}
\begin{split}
\bar{v}_t&=\int_0^tds \frac{e^{-\gamma_0(t-s)}}{m}f_s\times \\
&\times \left(\cos\left[(t-s)\Sigma\right]-\frac{\gamma_0}{\Sigma}\sin\left[\left(t-s\right)\Sigma\right]\right)\, 
\end{split}
\label{mean-velocity}
\end{equation}
and the fluctuating component by 
\begin{equation}
\begin{split}
v_t^\xi&=\int_0^tds \frac{e^{-\gamma_0(t-s)}}{m}\xi_s\times\\
&\times \left(\cos\left[(t-s)\Sigma\right]-\frac{\gamma_0}{\Sigma}\sin\left[\left(t-s\right)\Sigma\right]\right)\, .
\end{split}
\label{stoc-vel}
\end{equation}

\noindent For later purposes, we note that $\mathbb{E}[v_t]=\bar{v}_t$, $\mathbb{E}[v^{\xi}_t]=0$, 
\begin{equation*}
\mathbb{E}[v_t^2]=\left(\bar{v}_t\right)^2+\mathbb{E}\left[\left(v_t^{\xi}\right)^2\right]+2\bar{v}_t\, \mathbb{E}\left[v_t^{\xi}\right]\, ,
\end{equation*}
and
\begin{equation}
\lim_{t\to\infty}\mathbb{E}\left[\left(v_t^{\xi}\right)^2\right]=\frac{k_{\rm B}T}{ m}\, ,
\label{vsquared}
\end{equation}
where we have used the noise time-correlation function $\mathbb{E}\left[\xi_t\xi_s\right]=2\gamma k_{\rm B}T \delta(t-s)$. \\

\noindent {\it \bf Stochastic energetics of the mirror.} The power delivered by the engine can be determined from the stochastic energetics of the mirror.  \\
\noindent The total internal energy of the mirror is given by 
$$
U_t=\frac{m}{2}v_t^2+\frac{m\, \omega^2}{2}x_t^2\, ,
$$
and its derivative reads, using Eq.~\eqref{sys-first-order},
$$
\frac{d}{dt}U_t=-\gamma v_t^2+f_tv_t+v_t\xi_t\, .
$$
Taking the average over all realizations, and considering a time large enough so that we can neglect transient behaviour and use the result of Eq.~\eqref{vsquared} we obtain
\begin{equation}
\frac{d}{dt}\mathbb{E}\left[U_t\right]=\gamma \frac{k_{\rm B}T}{m}+f_t\bar{v}_t-\gamma \mathbb{E}\left[v_t^2\right]\, .
\label{en-balance}
\end{equation}

This equation shows that the variation of the energy balance of the mirror involves three components: the power delivered by the many-body quantum engine through the force $f_t$, the average power uptake from thermal fluctuations, which is proportional to the temperature $T$, and the energy that is dissipated by mechanical friction.

When the long-time dynamics of the mirror settles to an asymptotic cycle, we can compute the average power per cycle as 
$$
P_{\rm av}=\frac{1}{t_{\rm c}}\int_0^{t_{\rm c}} dt f_t v_t=\frac{\gamma}{t_{\rm c}} \int_0^{t_{\rm c}} dt \left(\mathbb{E}\left[v_t^2\right]-\frac{k_{\rm B}T}{m}\right) \, ,
$$ 
where $t_{\rm c}$ is  the period of the cycle. Using relation \eqref{vsquared}, this expression can be rewritten solely in terms of the mean velocity of the mirror
$$
P_{\rm av}=\frac{\gamma}{t_{\rm c}}\int_0^{t_{\rm c}} dt \, \bar{v}_t^2\, .
$$ 

When the force $f_t$ does not have an a priori known period, we can determine the delivered power through the long-time average 
$$
P_{\rm av}=\lim_{t_{\rm obs}\to\infty}\frac{\gamma}{t_{\rm obs}}\int_0^{t_{\rm obs}}dt \, \bar{v}_t^2\, .
$$ 
This formula for the average power is justified by the fact that the energy of the mirror remains bounded and thus the average energy variation goes to zero at long times,  
$$
\lim_{t_{\rm obs}\to\infty}\frac{1}{t_{\rm obs}}\left(\bar{U}_{t_{\rm obs}}-\bar{U}_0\right)=0\, .
$$
Hence, the energy absorbed by the mirror from the engine goes into an extra contribution of heat dissipated by friction, which adds to the equilibrium thermal one. \\

\noindent {\it \bf Numerical simulation of the engine dynamics and of the mirror. } In order to quantitatively analyse the performance of our engine, we need to devise a suitable numerical scheme. In practice, we want to determine a system of differential equations involving dimensionless variables which are representative of the engine-mirror dynamics. To this end, we focus on the mean contribution of the position and of the velocity, which determine the power output. 

\noindent We first define  the dimensionless time $\tau=\omega t$. Next we consider the deterministic force which drives the mirror dynamics; this term is given, in the finite-density limit, by Eq.~\eqref{force-fin-den}. By expressing it as a function of $\tau$, we obtain 
$$
f_{t=\tau\omega^{-1}}=\hbar \omega_{\rm cav}^0 D_0 F_\tau\, ,
$$
where we have extracted the dimensionless force
\begin{equation}
F_\tau=\frac{g}{\kappa}\left(\Gamma_0 s_+s_-\right)_\tau\, .
\label{dimless-force}
\end{equation}
This term depends on the atomic variables $s_{\pm}$ at the time $\tau=t\omega$, as well as on the value of $\Gamma_0$ at $\tau$, which is a function of the control parameter $\Delta$. The values of $s_\pm$ and $s_z$ are determined by the system of differential equations 
\begin{equation}
\begin{split}
\frac{d}{d\tau}s_+&=-i\frac{\Omega}{\omega} s_{z}-i\frac{\Delta}{\omega} s_+-i\frac{g}{\omega}C_0s_{ z}s_++\frac{g}{\omega}\frac{\Gamma_0}{2}s_{z}s_+\, ,\\
\frac{d}{d\tau}s_{z}&=2i\frac{\Omega}{\omega}\left(s_--s_+\right)-2\frac{g}{\omega}\Gamma_0s_+s_-\, .
\end{split}
\label{adim-sys-diff-atoms}
\end{equation}

\noindent The mean position of the mirror $\bar{x}_t$, which appears in Eq.~\eqref{mean-position}, can also be factorized into a dimensionless dynamical quantity and a dimensional constant depending on the details of the experimental setting,
\begin{equation}
\bar{x}_{t=\tau\, \omega^{-1}}= \frac{\omega_{\rm cav}^0}{\omega}\frac{\hbar D_0}{m \omega} X_\tau\, ,
\label{x-dim}
\end{equation}
where 
$$
X_\tau=\int_0^\tau dy \, \frac{\omega}{\Sigma}e^{-\frac{\gamma_0}{\omega}(\tau-y)}\sin\left[\left(\tau-y\right)\frac{\Sigma}{\omega}\right]F_y\, 
$$
and $\gamma_0=\gamma/(2m)$. Similarly, the mean velocity of the mirror \eqref{mean-velocity} can be expressed as
\begin{equation}
\bar{v}_{t=\tau \omega^{-1}}=\frac{\omega_{\rm cav}^0}{\omega}\frac{\hbar D_0}{m }V_\tau\, ,
\label{v-dim}
\end{equation}
where 
\begin{equation*}
\begin{split}
V_\tau &=\int_0^\tau dy \, e^{-\frac{\gamma_0}{\omega}(\tau-y)}F_y\times \\ &\times\left\{\cos\left[\left(\tau-y\right)\frac{\Sigma}{\omega}\right]-\frac{\gamma_0}{\Sigma}\sin\left[\left(\tau-y\right)\frac{\Sigma}{\omega}\right]\right\}\, .
\end{split}
\end{equation*}
The quantities $X_\tau,V_\tau$ are solution to the system of differential equations 
\begin{equation}
\frac{d}{d\tau}\begin{pmatrix}
X_\tau\\ V_\tau
\end{pmatrix}=\begin{pmatrix}
0&1\\
-1&-2\gamma_{0}/\omega
\end{pmatrix}\begin{pmatrix}
X_\tau\\ V_\tau
\end{pmatrix}+\begin{pmatrix}
0\\ F_\tau
\end{pmatrix}\, .
\label{adim-sys-diff-mirror}
\end{equation}

\noindent From  the equations \eqref{dimless-force}-\eqref{adim-sys-diff-mirror} the trajectory of the mirror can be determined numerically. The power that is delivered by the engine to the mirror can then be obtained from the relation 
\begin{equation}
\begin{split}
P_{\rm av}=&\frac{\gamma}{t_{\rm obs}}\int_0^{t_{\rm obs}}dt \, \bar{v}_t^{2}\\
=&\left(\frac{\omega_{\rm cav}^0}{\omega}\right)^2 \frac{\hbar^2 D_0^2\omega}{m} \left[2\gamma_0\frac{1}{\tau_{\rm obs}}\int_0^{\tau_{\rm obs}} d\tau \, V_\tau^2 \right]\, ,
\end{split}
\end{equation}
where $t_{\rm obs}$ is the observation time, which equals the period of the driving $t_{\rm obs}=2\pi/\omega$ for the periodic mode of operation and is taken to be very large for the time-crystal mode, $t_{\rm obs}\gg\omega^{-1}$.

\noindent Note that the rescaling of dynamical quantities keeps our analysis general and thus applicable to different experimental instances where $\omega_{\rm cav}^0,\omega$, the mass of the mirror and the linear density of atoms can vary.

\onecolumngrid
\newpage

\renewcommand\thesection{S\arabic{section}}
\renewcommand\theequation{S\arabic{equation}}
\renewcommand\thefigure{S\arabic{figure}}
\setcounter{equation}{0}
\setcounter{figure}{0}

\begin{center}
{\Large SUPPLEMENTARY INFORMATION}
\end{center}

\subsection{Effective Description and first-order expansion in $x/L_0$}

A reduced dynamical description of the ensemble of atoms can be obtained by adiabatically eliminating the cavity light field \cite{PhysRevA.56.2249,PhysRevLett.107.277201,Norcia259}. In this section, we review this procedure and discuss the resulting dynamical generator. \\

The Heisenberg equation of motion for the bosonic operator $a$ reads
$$
\dot{a}=(i\delta -\frac{\kappa}{2})a-i\frac{g}{\sqrt{N}}S_-\, ,
$$
and its formal solution is given by 
$$
a=e^{(i\delta -\frac{\kappa}{2})t}a_0-i\frac{g}{\sqrt{N}}\int_0^tdu \, e^{(i\delta -\frac{\kappa}{2})(t-u)}(S_-)_u\, .
$$
In the adiabatic limit, where the operators $S_\pm$ in the Heisenberg picture are practically constant on the relaxation time scale $1/\kappa$, this expression can be simplified to 
$$
a\approx -i\frac{g}{\sqrt{N}}S_-\int_0^tdu \, e^{(i\delta -\frac{\kappa}{2})(t-u)}\approx \frac{-2igS_-}{\sqrt{N}(\kappa-2i\delta)}\, ,
$$
where $S_- = (S_-)_{u=t}$. Hence, the bosonic operator can effectively be replaced with the rescaled collective spin operator 
\begin{equation}
a=\frac{2g}{\sqrt{N}(2\delta +i\kappa)}S_-\, .
\label{slaving}
\end{equation}
Substituting this result into the total Hamiltonian of the system,
$$
H_{\rm tot}=H_{\rm L}+H_{\rm int}+H_{\rm ph}\, ,
$$
yields the effective Hamiltonian 
$$
\tilde{H}=H_{\rm L}+\frac{g \hbar }{N}\, C\, S_+S_-\, , \qquad \mbox{with}\qquad C=\frac{4\,g\delta}{4\delta^2+\kappa^2}\, .
$$
For the effective dissipator we obtain 
$$
\tilde{\mathcal{D}}[\rho]=\frac{g\hbar}{N}\Gamma\left( S_-\rho\,  S_+-\frac{1}{2}\left\{S_+S_-,\rho\right\}\right)\, , \qquad \mbox{with}\qquad \Gamma=\frac{4\, g\kappa}{4\delta^2+\kappa^2}\, .
$$
\\

Having derived an effective model for the ensemble of atoms, we now have to account for small oscillations of the mirror around its equilibrium position. Changing the length of the cavity alters the wave-length, and thus the frequency $\omega_{\rm cav}=nc/(2L)$, of the light mode, where $c$ denotes the speed of light and $n$ is a positive integer. Hence the parameter $\delta = \omega_{{{\rm at}}} + \Delta -\omega_{{{\rm cav}}}$, which enters the effective Hamiltonian and the effective dissipation super operator, becomes a function of the the length $L= L_0 + x$. 

For small oscillations, i.e., $x/L_0\ll 1$, we have
$$
\omega_{\rm cav}=\frac{n\, c}{2(L_0+x)}\approx \omega_{\rm cav}^0\left[1-\frac{x}{L_0}\right]\, \quad \text{with}\quad 
\omega^0_{{{\rm cav}}} = \frac{nc}{2L_0},
$$
and  
$$
\delta\approx\delta_0+\omega_{\rm cav}^0\frac{x}{L_0}\, \quad \text{with}\quad \delta_0=\omega_{\rm at}+\Delta-\omega_{\rm cav}^0\, .
$$
Consequently, the dimensionless constants $\Gamma$ and $C$ become
$$
\Gamma\approx \Gamma_0-\Gamma_1\frac{x}{L_0}\, ,\qquad \mbox{with} \qquad \Gamma_0=\frac{4\, g\kappa}{\kappa^2+4\delta_0^2}\, ,\qquad \mbox { and } \qquad \Gamma_1=\omega_{\rm cav}^0\Gamma_0\frac{8\, \delta_0}{\left(4\delta_0^2+\kappa^2\right)}\, ,
$$
and 
$$
C\approx C_0-C_1\frac{x}{L_0}\, ,\qquad \mbox{with} \qquad C_0=\frac{4\, g\delta_0}{\kappa^2+4\delta_0^2}\, ,\qquad \mbox { and } \qquad C_1=\frac{\omega_{\rm cav}^0}{\delta_0}C_0\frac{\left(4\delta_0^2-\kappa^2\right)}{\left(4\delta_0^2+\kappa^2\right)}\, .
$$

\noindent These relations show how the reduced dynamical description of the atoms depends on the displacement of the mirror $x$. 

\noindent By following the same lines, we find that the force acting on the mirror is given by 
$$
f_t=-\langle \left[\frac{\partial}{\partial x}H_{\rm ph}^{\rm S} \right]\rangle_t\approx \frac{\omega_{\rm cav}^0}{L_0}\braket{ a^\dagger a}_t\approx \frac{\hbar g}{N}\frac{\omega_{\rm cav}^0}{\kappa \, L_0}\left(\Gamma_0-\Gamma_1\frac{x}{L_0}\right) \braket{S_+S_-}_t\, .
$$ 
where we have used the relation  \eqref{slaving} to eliminate the operators pertaining to the light field.

\subsection{Dynamics in the finite density of atoms case.}
In this section, we provide further details on how the non-linear differential equations, which govern the dynamics of the collective variables $s_\pm$ and $s_z$, can be obtained from the effective description of the atomic system in the limit of finite density. 

The Heisenberg equation of motion for the rescaled operator $S_+/N$ is given by
\begin{equation}
\frac{d}{dt}\frac{S_+}{N}=i\left[H_{\rm eff},\frac{S_+}{N}\right]+\frac{g_0}{N}\frac{\Gamma_0}{2}\left(\left[S_+,\frac{S_+}{N}\right]S_-+S_+\left[\frac{S_+}{N},S_-\right]\right)=-i\Omega \frac{S_{ z}}{N}-i\Delta \frac{S_+}{N}-i\frac{g}{N}C_0\frac{S_{ z}}{N}\frac{S_+}{N}+\frac{g\Gamma_0}{2}\frac{S_{ z}}{N}\frac{S_+}{N}\, .
\label{eq-mf}
\end{equation}
We now take the average of both sides of this differential equation with respect to the initial state $\rho_0$ of the atomic system and observe that quadratic contributions can be factorized as 
\begin{equation}
\braket{\frac{S_{ z}}{N}\frac{S_+}{N}}\longrightarrow \braket{\frac{S_{ z}}{N}}\braket{\frac{S_+}{N}}
\label{fact-quad-av}
\end{equation}
for $N\gg 1$, since the coupling between individual constituents of the ensemble is of order $1/N$. This approximation becomes exact in the limit $N\to\infty$ \cite{Benatti_2018}. Furthermore, this scheme naturally extends to breaking up the time-evolution into a series of intervals with constant dynamical generator, which is the case for the quenched protocol adopted for the detuning $\Delta$. 

\noindent Using the relation \eqref{fact-quad-av} and defining 
$$
s_\pm=\lim_{N\to\infty}\braket{\frac{S_\pm}{N}}\, ,\qquad s_{ z}=\lim_{N\to\infty}\braket{\frac{S_{ z}}{N}}\, ,
$$
yields the non-linear differential equation
$$
\dot{s}_+=-i\Omega s_{ z}-i\Delta s_+-igC_0s_{ z}s_++\frac{g\Gamma_0}{2}s_{ z}s_+\, .
$$
Along the same lines, we find the equation of motion for $s_{z}$,
$$
\dot{s}_{ z}=2i\Omega\left(s_--s_+\right)-2g\Gamma_0s_+s_-\, .
$$

\end{document}